\documentclass[manuscript]{aastex6}
\usepackage{epsfig}
\usepackage{graphicx}
\usepackage{amssymb}
\usepackage{amsmath}
\usepackage{times}
\usepackage{hyperref}
\usepackage{natbib}
\usepackage{blindtext}
\usepackage{roussev}
\usepackage{color}
\usepackage{mathrsfs}

\newcount\doepsf
\newcommand{\alf}{Alfv\'en }

\newcommand{\be}{\begin{equation}}
\newcommand{\ee}{\end{equation}}
\newcommand{\bea}{\begin{eqnarray}}
\newcommand{\eea}{\end{eqnarray}}
\newcommand{\bean}{\begin{eqnarray*}}
\newcommand{\eean}{\end{eqnarray*}}

\begin{document}

\title{Toward  Quantitative Model for Simulation and Forecast of Solar Energetic Particles Production during Gradual Events - I: Magnetohydrodynamic Background Coupled to the SEP Model}

\author{
D. Borovikov,     \altaffilmark{1,2}
I.~V. Sokolov,    \altaffilmark{1} 
I.~I. Roussev,    \altaffilmark{3}
A.~Taktakishvili, \altaffilmark{4,5}
and T.~I. Gombosi,\altaffilmark{1}
}

\altaffiltext{1}{Center for Space Environment Modeling, University of Michigan,
2455 Hayward St, Ann Arbor, MI 48109; \\dborovik@umich.edu, igorsok@umich.edu,
tamas@umich.edu.}
\altaffiltext{2}{Space Science Center, University of New Hampshire, 8 College Road Durham, NH 03824}
\altaffiltext{3}{National Science Foundation; iroussev@nsf.gov}
\altaffiltext{4}{Community Coordinated Modeling Center, NASA Goddard Space Flight Center, Greenbelt, MD 20771, USA; Aleksandre.Taktakishvili-1@nasa.gov}
\altaffiltext{5}{Catholic University of America, Washington, DC 20064, USA}

\begin{abstract}
Solar Energetic Particles (SEPs) are an important aspect of space weather.
SEP events posses a high destructive potential, 
since they may cause disruptions of communication systems on Earth
and be fatal to crew members onboard spacecrafts and, 
in extreme cases, harmful to people onboard high altitude flights.
However, currently the research community lacks efficient tools to
predict such hazardous threat and its potential impacts.
Such a tool is a first step for mankind to improve its preparedness for
SEP events and ultimately to be able to mitigate their effects.
The main goal of the presented research effort
is to develop a computational tool that
will have the forecasting capability and can be serve in
operational system that will provide live information on the current potential
threats posed by SEP based on the observations of the Sun.
In the present paper the fundamentals of magneto-hydrodynamical (MHD)
simulations are discussed to be employed as 
a critical part of the desired forecasting system.
\end{abstract}


\numberwithin{equation}{section}

\section{Introduction}
\label{sec:intro}
\subsection{Potential threats of SEP}
For our technologically advanced civilization, space plays an increasingly
important role.
The idea of interplanetary travel and even establishing
colonies on Moon and Mars slowly but steadily 
transitions from science fiction into the realm of plausibility.
However fascinating as the perspectives could sound,
our ability to predict dangers that we may encounter along the way
needs to be significantly improved.
The dangers themselves, however, are not unknown to us.
One of them comes from our Sun 
in the form of Solar Energetic Particles (SEP).
Triggered by extreme solar events, SEP fluxes may reach values that
are damaging to the electronics onboard spacecraft
and potentially fatal to the crews \citep[e.g.][]{joyce15}.
Precedents of SEP events of such ominous scale have been recorded 
in the recent history.

During the Apollo program when astronauts repeatedly visited the Moon, 
a huge SEP event accompanied the major August 1972 solar storm. 
The integrated SEP flux produced by this
storm could have been fatal for Moon walking astronauts since 
the radiation dose from energetic particles
penetrating their spacesuits would have exceeded the lethal level 
(${\sim}400$ rems in a short period of time).
Luckily, during this event the Apollo~16 astronauts were already safely 
back on the Earth, while the crew of
Apollo 17 was still preparing for their mission. 
An SEP event during the historic
``giant leap for mankind'' lunar landing could have been fatal. 
NASA, and the entire world, was lucky that
the Sun ``cooperated'' with this endeavor.

However, when planning interplanetary human missions, one cannot rely on luck. 
A mission to Mars and
back will take several years,
 and there is a significant risk to have one or more extreme SEP events 
that ``may
expose the crew to doses that lead to acute radiation effects.''
The fact that we cannot predict SEP events
makes a human mission to Mars a ``high-risk adventure"
\citep{Hellweg:2007,Jakel:2004}.

Let us come back to the Earth and consider the harmful effects of SEP 
events on assets at Low Earth Orbit (LEO). 
The terrestrial magnetic field  provides
some shielding for the International Space
Station (ISS) as well as the majority of unmanned missions from SEPs. 
However, extreme SEP events, such as that of 20~January~2005 
\citep[e.g.][]{grechnev08,matthia09}, 
have hard energy spectra and they are particularly rich in hundreds of MeV to
several GeV protons. 
A significant fraction of flux of the high-energy particles,
which have a high penetrating capability, can reach 
LEO, 
thus producing significant radiation hazard for human spaceflight. 
Comparing with the direct threat to human life and health, 
the SEP effect on unmanned satellites on LEO may seem to
be not so important. 
However, the possible loss of entire satellites with their expensive computers,
sensors, and other elements of electronics is not limited to the cost 
(typically hundreds of millions of dollars) of the satellite itself. 
Many satellites are integrated into vitally important systems of defense, 
rescue, navigation, so the disruption of such system may have 
catastrophic consequences.

Even closer to the Earth is the ozone (O$_3$) layer in the stratosphere. 
This layer protects the Earth against harmful solar UV and EUV emissions, 
and the depletion of the ozone layer would increase the number of skin
cancer cases in the human population. 
The higher energy SEPs can reach the stratosphere. 
In particular, the SEP event in August 1972 reduced the ozone 
concentration near the North geomagnetic pole by $>20\%$, 
this reduction lasted for ${\sim}20$ days, 
i.e. well after the end of the SEP event \citep{Heath77}. 
The reason is that very large SEP events can increase the ionization degree
in the stratosphere by more than a factor of 100 over
quiet times \citep[see][]{Makhmutov:2009}. 
The increased ionization initiates a chain of chemical reactions
that produce so-called ``odd nitrogen'' molecules, 
like NO, which cannot be created from even nitrogen molecule, N$_2$. 
Molecules like NO catalyze the ozone decay, 
and each odd nitrogen molecule can ``kill'' millions of O$_3$ molecules.

Among other threats, SEP events and their increased radiation hazard make 
the flight routes over the North Pole more challenging, because of
the increased risk of radiation exposure and interference with 
communication in high frequency (HF) range \citep{Morris07}. 

We see that some effects of extreme SEP events are only important 
at higher latitudes near the geomagnetic poles, 
approximately in the same regions where auroras are often observed. 
However, during relatively infrequent but more powerful events, 
such as the Carrington event of 1859, the aurora had been observed
as far from geomagnetic poles as at Hawaii, Miami or Jerusalem 
\citep{Cliver06,Green06,Green06a,Shea06,Shea06a}. 
For such events, the area in which the ozone
layer is depleted may also extend well beyond the polar region and 
this would last longer. 
Air traffic may be interrupted all over the world. 
We know that such unique events may happen, but we do not know, how
it would affect the modern technology. 

These are the main reasons why SEP events are considered as one 
of the most important aspects of space weather.
This explains the need in a predictive technology
that is capable of providing a reliable quantitative forecast of SEP
events and their impacts.
\subsection{Mechanisms of SEP Production}

First observations of ``solar cosmic rays'',
as SEP were referred to at the time,
dated back to 1942 \citep[see][]{forbush46} and were immediately linked
to solar flares that preceded these particles events.
This hypothesis was further supported by observations that
followed, and solar flares were considered to be the primary source
of SEPs \citep{meyer56}.
As number of observed events increased, 
it became apparent that features of events,
such as the aforementioned composition, duration, etc.,
exhibit a wide variability \citep{wild63}.
The discovery of Coronal Mass Ejections (CMEs)
prompted formulation of a new hypothesis
that SEPs are produced by interplanetary shocks that often
accompany flares rather than by flares themselves
\citep{kahler78,Gosling1993}.

The debate ultimately resulted in 
the commonly adopted paradigm 
\citep[e.g.][]{Ruffolo1998,reames99,Reames2002,cliver09} 
that states that SEP events 
can be divided into two distinct classes:
(1)~impulsive, and (2)~gradual events.
The former are caused by solar flares, 
while the latter are associated with CMEs.
Gradual events are prolonged in time and 
more extended in longitudinal range compared to the impulsive events.
Also, 
SEP composition was found to be a good indicator of nature of events 
\citep{cane06}, e.g. flare associated events have Fe/O abundance of~${\sim}1$
and are electron-rich, 
while those associated with CME-driven shocks have Fe/O abundance of~
${\sim}0.1$ and are proton-rich.

It should be noted that the pattern above was originally discovered for 
particles of energies that are limited by ${\sim}$30~MeV.
Such particles are easily detected by instruments outside 
the Earth's magnetosphere.
For this reason, 
many models focus on this particular part of the particle
population.
However, the particles that pose the largest threat are those that
exceed this energy, and for them the aforementioned pattern is much less clear.
Based on compositional and other data,
many large SEP events show signatures of both gradual and impulsive events and,
thus, don't fully agree with simple bi-modal paradigm 
\cite[e.g.][]{cohen99,mazur99}.
\citet{cohen08} explain the discrepancy by simultaneous
particle producion both at flare sites and CME shock fronts,
while \citet{tylka05} 
suggest that there is no true separation of events into two
distinct categories:
seed suprathermal particle population originates in flares 
and is then accelerated by CME-driven interplanetary shocks.
In both of these explanations, signatures of both types of events naturally
arise in large SEP events.
These events are frequently
associated with so-called Ground Level Events (GLEs),
as most energetic particles have the potential to 
penetrate Earth's magnetosphere and ionosphere
\citep{Shea:1990,shea94,shea12,gopalswamy14}.
Relevant data are provided by measurements performed with neutron monitors.
Analysis of properties of GLEs and those of
associated solar flares and CMEs 
\citep[e.g.][]{kahler12,gopalswamy12}
confirms that SEP production is likely to involve both types of events.
Further observations, e.g. by AMS-02 instrument \citep{PRLprotons},
will provide valuable insights into this problem.
In our work, we focus primarily on gradual, shock-driven events.

Solar eruptions, including CMEs, 
are associated with a major restructuring of the
coronal magnetic field and the ejection of solar material 
($\sim$~$10^{12\div13}$~kg) and magnetic flux ($\sim$~$10^{13\div15}$~Wb) 
into interplanetary space \citep[e.g.][]{Roussev2006}.
A shock wave driven by the ejecta can accelerate 
charged particles to ultra-relativistic energies as the result of 
Fermi acceleration processes \citep{fermi49}.
 The diffusive-shock-acceleration (DSA) is a profound mechanism, 
which naturally produces the observed power-law spectra of energetic particles.
 It was first proposed by 
\cite{Krymsky1977,Axford1977,Bell1978a,Bell1978b,Blandford1978} 
to explain an origin of galactic cosmic rays, however, 
for the past four decades, 
this mechanism has been studied extensively also in the context of co-rotating 
and traveling interplanetary shocks and has been demonstrated to be well 
supported both by theory 
\citep[e.g.][]{Lee1997,Ng1999,Ng2003,Zank2000} and observations 
\cite[e.g.][]{Kahler1994,Tylka1999,Cliver2004,tylka05}.

The most efficient particle acceleration takes place near the Sun 
at heliocentric distances of $2\div15\, R_{\odot}$, 
and the fastest particles can escape upstream of the shock, 
then propagating along the lines of the interplanetary magnetic field 
and reaching the Earth shortly after the initiation of the CME ($\le 1$~hr).  

The theory of DSA is being debated within the community \cite[]{reames99,Reames2002, Tylka2001}, 
since very little is known from observations about the dynamical properties of CME-driven shock waves in the inner corona soon 
after the onset of an eruption.
 The main argument against the shock origin is that 
near the Sun the ambient \alf speed is so large, 
due to the strong magnetic fields there, 
that a strongly super-magnetosonic shock wave
is difficult to anticipate \citep{Gopalswamy2001}.  
How soon after the onset of a CME the shock wave forms, 
and how it evolves in time depends largely on how this shock
wave is driven by the erupting coronal magnetic fields.  

To address the issue of shock origin during CMEs, it is required
that real magnetic data are incorporated into a
global model of the solar corona, as this had been done in,
for example, \citet{Roussev2004}.  
As proposed by \citet{tylka05}, 
the shock geometry plays a significant role in the spectral and
compositional variability of SEPs above ${\sim}30$ MeV/nuc.  
Therefore, in order to explain the observed signatures of gradual SEP events, global models of solar eruptions are required to explain 
the time-dependent changes in the strength and geometry of shocks 
during these events.
 The CME-driven shock continues to accelerate particles, and the shock passage at 1 AU is often accompanied by an enhancement of the energetic-particle flux.
  To simulate this effect, 
the shock wave evolution should be continuously traced while it 
propagates to 1 AU.

\subsection{Goal and content of the paper}
The goal of our current and future research
is to develop the computational framework embracing several coupled physical 
and numerical models, 
which could 
quantitatively simulate the SEP production during the gradual events with ultimately 
achieving a capability to predict the SEP flux and spectrum 
(or, at least, the probability 
of dangerously high flux and related radiation hazard). 

In the series of two papers 
we  outline  the framework and describe its  existing components.
The present paper is the first in the series,
it contains the review of the MHD component of the framework.
The second paper will describe the kinetic component of the framework.

Some of the computational models or their couplings are still
underdeveloped.
Therefore, the paper mostly focuses on presenting the physical and
mathematical fundamentals of the integrated model,
as well as on the description of the available computational tools
and their integration into the framework, rather than on the particular
results of the models.
Some numerical results here are provided only to illustrate 
operation of the models.
From the brief discussion above we can summarize, which
computational tools and technologies are needed to achieve 
the claimed research goal, therefore, what should be included into the
desired computational framework.

First, 
one needs to simulate a full 3-D structure of the interplanetary magnetic field prior to CME, which 
determines the magnetic connectivity and allows simulating the SEP transport along the 
magnetic field lines toward 1 AU.
 One also needs  to know the 3-D distribution of the solar wind 
parameters.
 This ambient solution affects the CME and shock wave travel time to 1 AU, hence, 
the time of SEP enhancement in the course of the shock wave passage.
The pre-eruptive structure of the Solar Corona (SC)
should also be known, 
since it controls the possibility for the shock wave formation at small 
heliocentric distances, which results in efficient DSA, 
as well as the magnetic connectivity of the Active 
Region (AR), at which CME originates, to the upper  SC.
To simulate the ambient solution in the SC and inner heliosphere (IH),
the \alf wave turbulence based Solar atmosphere Model (AWSoM) is used 
as described in this paper in Section~\ref{sec:MHD:steady}.
 In order to simulate an ongoing CME and to reach the predictive capability, 
the model should run faster than the Real-time, therefore, 
the AWSoM-R model with this feature is utilized
(presented in Section~\ref{sec:MHD:AWSOMR}). 
The CME-driven shock wave should be simulated starting from the lower altitude.
Relevant models are described in Section~\ref{sec:MHD:CME}.

{The overview of MHD models and tools presented in this paper wouldn't
be complete without demonstrating how they fit into the overall framework
and what makes them an irreplaceable piece in the puzzle.
This requires a summary, however brief,
of a particle code utilized in simulations (Section~\ref{sec:kinetic}).}
{
We conclude the paper with the proof-of-concept results
obtained with the help of our newly developed SEP forecasting framework
(Section~\ref{sec:Results}).}
Again,
{a detailed summary of} physical models and numerical tools that will be used to describe the kinetic
component of the framework are to be considered in the second paper 
of the series.
{
\section{Alfv{\'e}n Wave Turbulence Driven MHD Description of the Solar Corona and the Solar Wind
}
\label{sec:mhd}

\subsection{Steady-state solar corona and solar wind}
\label{sec:MHD:steady}
In any predictive model for eruptive solar events,
the background steady-state SC and IH
are as important as a stage for a performance.
If poorly designed, the foundation would compromise the whole facility.
Thus, an accurate and carefully validated model
for the steady-state background is vital
and shouldn't be overlooked or explored superficially.
In our work we use a widely accepted paradigm
that the solar wind is driven by  and the SC is heated by the dissipation in,
the \alf wave turbulence.
\subsubsection{Alfv{\'e}n wave turbulence}
The concept of \alf waves was introduced more than 70 years ago
by \citet{Alfven1942}.
 The importance of the role they play within the Solar system
was not immediately recognized due to the lack of relevant observations.
 Results from Mariner 2 allowed
a data-backed study of a wave-related phenomena in solar wind.
 A detailed analysis of these observations can
be found in, for example, \citet{Coleman1966,Coleman1967}.
 This pioneering study culminated in \citet{Coleman1968},
a work that stated that \alf wave turbulence has the potential to drive solar wind  in a way that is consistent with observations at 1 AU.

Attention to \alf waves related phenomena was continuously increasing and
an ever growing number of studies on interaction of these waves  with 
solar wind plasma and 
various aspects of associated effects were undertaken.
Examples of the earliest efforts to investigate the role of \alf waves in 
solar wind acceleration 
are \citet{Belcher1969,Belcher1971,alazraki71}.
A consistent and comprehensive theoretical description of
\alf wave turbulence and its effect on the averaged plasma motion
has been developed in a series of works,
particularly, \citet{Dewar1970} and \citet{Jacques1977,Jacques1978}
(see also references therein).
More recent efforts to simulate solar wind acceleration utilize the approach
developed in these works \citep[e.g.][]{Usmanov2000}.
Currently, it is commonly accepted, that the gradient of 
the \alf wave pressure
is the key driver for the solar wind acceleration.

At the same time, 
damping of \alf wave turbulence as a 
source of the coronal heating
was extensively studied \citep[e.g.][]{Barnes1966,Barnes1968}.
Later, 
it was demonstrated that reflection from the sharp pressure gradients 
in the solar wind \citep{Heinemann1980,Leroy1980}
is a critical component of \alf wave turbulence damping 
\citep{Matthaeus1999,Dmitruk2002,Verdini2007}.
For this reason, 
many numerical models explore
the generation of reflected counter-propagating waves
as the underlying cause of the turbulence energy cascade 
\citep[e.g.][]{Cranmer2010},
which transports the energy of turbulence from the large scale motions
across the {\it inertial range} of the turbulence spatial scale
to short-wavelength perturbations.
The latter can efficiently damp due to the wave-particle interaction.
In this way,
the turbulence energy is converted to the particle (thermal) energy.

Recent efforts of many studies are aimed at developing
models that include \alf waves as a primary driving agent for both heating and 
accelerating of the solar wind.
Examples are 
\citet{Hu2003,Suzuki2005,Verdini2010,Matsumoto2012,
Lionello2014a,Lionello2014b}.

\subsubsection{Ad Hoc Coronal Heating Functions and Semi-Empirical Models for the Solar Wind Heating}
It is important to emphasize, 
that while incorporating the \alf wave driven acceleration 
is a matter of including
the wave pressure gradient into governing equations \citep{Jacques1977},
there is still no widely accepted approach to describing 
the coronal heating 
via \alf wave turbulence cascade.
A large number of models of SC heating
have been constructed over the years.
One can trace two major approaches to representing the process:
(i) to use an ad-hoc heating function to mimic SC heating with
heating rate being chosen to better fit observations;
(ii) to use a semi-empirical coronal heating function that is based
on aspects of physics of \alf waves.

The former approach is utilized, for example,
by \citet{Lionello2001,Lionello2009,Riley2006,Titov2008,Downs2010}.
This method provides a reasonably good agreement with observations in EUV,
X-rays and white light.
The agreement looks particularly impressive for the PSI predictions
about the solar eclipse image \citep{Mikic2007}.
An important limitation is that
models utilizing an ad-hoc approach depend on a few
free parameters, which need to be determined for various solar conditions.
Such approach has an inherent shortcoming: although it is well-suited
for typical conditions, it can't properly account
for unique conditions as those that can take place during extreme solar events.

Another illustration of the {\it ad hoc} approach is the
semi-empirical model to simulate solar wind.
For example,
the Wang-Sheeley-Arge (WSA) model
instead of incorporating physical properties of \alf waves,
utilizes semi-empirical formulae that relate the solar wind speed
with the solar magnetogram and the properties of the magnetic
field lines of the potential magnetic field as recovered
from the synoptic magnetogram.
Its development history may be traced through
\citet{Wang1990,Wang1992,Wang1995,Arge2000,Arge2003}.
The major benefit of the model is the opportunity to
seamlessly integrate it into a global space weather simulation
as was done in \citet{cohen07}.
In this study, the WSA formulae were used as the boundary condition
for the MHD simulator via the varied polytropic gas index distribution
\citep[see][]{Roussev2003}.
Models mentioned above successfully explain observations
of the solar wind parameters at 1 AU.

A number of validation and comparison studies have been published
\citep{Owens2008,Vasquez2008,MacNeice2009,Norquist2010,Gressl2014,
Jian2015,Reiss2016}.

However,
these models don't fully capture the physics of \alf wave turbulence
or even disregard it altogether.
Even though some models are designed to account for the \alf waves' physics
\citet[such as][]{cohen07},
neither does capture every aspect of the interaction of
the turbulence with the background flow, which include both energy
and momentum transfer from the turbulence
to the solar wind plasma.
Thus, neither model can be used
as a fully consistent tool for simulating the solar atmosphere.
\subsubsection{Alfv{\'e}n-Wave-Turbulence-Based Model for the Solar Atmosphere}
\label{sec:MHD:AWSOM}
 The {\it ad hoc} elements
were eliminated from the model for the SC and quiet-time IH
 by \citet{Sokolov2013}.
In the Alfv\'en Wave turbulence based Solar atmosphere Model (AWSoM) 
the plasma is heated
by the dissipation of the Alfv\'en wave turbulence, which, in turn,
is generated by the nonlinear interaction between oppositely propagating
waves \citep[]{Hollweg1986}.
Within the coronal holes, there are no closed magnetic field lines, hence,
there are no oppositely propagating waves.
Instead, a weak reflection of the outward propagating waves locally
generates sunward propagating waves as quantified by \citet{Holst2014}.
The small power in these locally generated
(and almost immediately dissipated) inward propagating waves leads
to a reduced turbulence dissipation rate in coronal holes,
naturally resulting in the bimodal solar wind structure.
Another consequence is that coronal holes look like cold black spots
in the EUV and X-rays images, 
while the closed field regions are  hot and bright,
and the brightest are active regions,
near which the wave reflection is particularly strong
\citep[see][]{Sokolov2013, Oran2013, Holst2014}.

The model equations are the following:
\begin{equation}\label{eq:cont}
\frac{\partial\rho}{\partial t}+\nabla\cdot(\rho{\bf u})=0,
\end{equation}
\begin{equation}\label{eq:induction}
\frac{\partial{\bf B}}{\partial t}+\nabla\cdot\left({\bf u}{\bf
  B}-{\bf B}{\bf u}\right)=0,
\end{equation}
\begin{equation}\label{eq:momentum}
\frac{\partial(\rho{\bf u})}{\partial t}+\nabla\cdot\left(\rho{\bf
  u}{\bf u}-\frac{{\bf B}{\bf
    B}}{\mu_0}\right)+\nabla\left(P_i+P_e+\frac{B^2}{2\mu_0}+P_A\right)=-\frac{GM_\sun\rho{\bf
    R}}{R^3},
\end{equation}
The notation used in the equations is as follows:
$\rho$ is the mass density,
$\mathbf{u}$ is the velocity, $u=|{\bf u}|$,
assumed to be the same for the ions and electrons,
$\mathbf{B}$ is the magnetic field, $B=|{\bf B}|$,
$G$ is the gravitational constant,
$M_\odot$ is the solar mass,
$\mathbf{r}$ is the position vector relative to the center of the Sun,
$R = |{\bf R}|$,
$\mu_0$ is the magnetic permeability of vacuum.
As has been shown by \citet{Jacques1977},
the \alf waves exert an {\it isotropic} pressure
(see term $\nabla P_A$ in the momentum equation).
The relation between the wave pressure and wave energy density is
$P_A=(w_++w_-)/2$.
 Herewith, $w_\pm$ are the energy densities for the turbulent waves
propagating along the magnetic field vector ($w_+$) or
in the opposite direction ($w_-$).
 The isotropic ion pressure, $P_i$, and electron pressure, $P_e$,
are governed by the energy equations:
\begin{eqnarray}\label{eq:energy}
\frac{\partial}{\partial
  t}\left(\frac{P_i}{\gamma-1}+\frac{\rho u^2}2+\frac{{\bf B}^2}{2\mu_0}\right)+\nabla\cdot\left\{\left(\frac{\rho u^2}2+\frac{\gamma P_i}{\gamma-1}+\frac{B^2}{\mu_0}\right){\bf
  u}-\frac{{\bf B}({\bf u}\cdot{\bf B})}{\mu_0}\right\}=\nonumber\\
= -({\bf u}\cdot\nabla)\left(P_e+P_A\right)+
\frac{N_eN_ik_B}{\gamma-1}\left(\frac{\nu_{ei}}{N_i}\right)\left(T_e-T_i\right)-\frac{GM_\sun\rho{\bf R}\cdot{\bf u}}{R^3}+Q_i,
\end{eqnarray}
\begin{eqnarray}\label{eq:electron}
\frac{\partial}{\partial
  t}\left(\frac{P_e}{\gamma-1}\right)&+&\nabla\cdot\left(\frac{P_e}{\gamma-1}{\bf
  u}\right)+P_e\nabla\cdot{\bf u}=\nonumber\\
&=&-\nabla\cdot\mathbf{q}_e
+\frac{N_eN_ik_B}{\gamma-1}\left(\frac{\nu_{ei}}{N_i}\right)\left(T_i-T_e\right)-Q_{\rm rad}+Q_e,
\end{eqnarray}
where
$T_{e,i}$ are the electron and ion temperatures,
$N_{e,i}$ are the electron and ion number densities,
and $k_B$ is the Boltzmann constant.
Other newly introduced terms are explained below.

The equation of state $
P_{e,i}=N_{e,i}k_BT_{e,i},
$
is used for both species.
 The polytropic index is $\gamma=5/3$.
The optically thin radiative energy loss rate in the lower corona is given by
\begin{equation}
Q_{\rm rad}=N_eN_i\Lambda(T_e)
\end{equation}
where
$\Lambda(T_e)$ is the radiative cooling curve taken from the
CHIANTI version 7.1 database \citep[and references therein]{landi13}.
The Coulomb collisional energy exchange rate between ions and electrons
is defined in terms of
the collision frequency
\begin{equation}
\frac{\nu_{ei}}{N_i}= \frac{2\sqrt{m_e}L_{C}
        (e^2/\varepsilon_0)^2
}{
         3 m_p(2\pi k_BT_e)^{3/2}}
\end{equation}
The electron heat flux $\mathbf{q}_e$ is used in
the collisional formulation of \citet{spitzer53}:
\begin{equation}
\mathbf{q}_e = \kappa_\|\mathbf{bb}\cdot\nabla T_e,\quad
\kappa_\|=3.2\frac{6\pi}{\Lambda_C}\sqrt{\frac{2\pi}{m_e}\frac{\varepsilon_0}{e^2}^2}
\left(k_BT_e\right)^{5/2}k_B
\end{equation}
where
$m_e$ and $e$ are the electron mass and charge,
$m_p$ is the proton mass,
$\varepsilon_0$ is the vacuum permittivity,
$\mathbf{b}=\mathbf{B}/B$,
$\Lambda_C$ is the Coulomb logarithm.

Dynamics of \alf wave turbulence and its interaction with 
the background plasma requires a special consideration. 
The evolution of the \alf wave amplitude
(velocity, $\delta\mathbf{u}$, and magnetic field, $\delta\mathbf{B}$)
is usually treated in terms of the \citet{Elsasser1950}
variables, $\mathbf{z}_\pm=\delta\mathbf{u}\mp\frac{\delta\mathbf{B}}{\sqrt{\mu_0\rho}}$.
The Wentzel-Kramers-Brillouin (WKB) approximation is used 
to derive the equations that govern transport of \alf waves, 
which may be reformulated in terms of the wave energy densities, 
$w_\pm=\rho\mathbf{z}^2_\pm/4$.
Dissipation of \alf waves, $\Gamma_\pm w_\pm$, 
is crucial in driving the solar wind and heating the coronal plasma.
The dissipation occurs, when two counter-propagating waves interact. 
Therefore, an efficient source of both types of waves is needed and 
it is maintained by \alf wave reflection from steep density gradients.
 For this reason, we need to go beyond the WKB approximation, which assumes 
that wavelength is much smaller than spatial scales in the background.
The equation describing propagation of the turbulence, its dissipation and 
reflection has been derived in \citet{Holst2014}:
\begin{equation}
\label{eq:w_pm}
\frac{\partial w_\pm}{\partial t} + \nabla\cdot\left[
(\mathbf{u}\pm\mathbf{V}_A)w_\pm
\right]
+\frac{w_\pm}{2}\left(\nabla\cdot\mathbf{u}\right) =
-\Gamma_\pm w_\pm\mp\mathcal{R}\sqrt{w_-w_+}.
\end{equation}
Here, the dissipation rate equals 
$\Gamma_\pm = \frac{2}{L_\perp}\sqrt{w_\mp/\rho}$
and the reflection coefficient is given by
\bea
{\cal R}=\min\left\{\sqrt{\left({\bf b}\cdot[\nabla\times{\bf
    u}]\right)^2+\left[({\bf V}_A\cdot\nabla)\log
    V_A\right]^2},\max(\Gamma_\pm)\right\}\times\nonumber\\
\times\left[\max\left(1-\frac{I_{\rm max}}{\sqrt{{w_+}/{w_-}}},0\right)-\max\left(1-\frac{I_{\rm max}}{\sqrt{{w_-}/{w_+}}},0\right)\right],
\eea
where $I_{\rm max}=2$  is the maximum degree of the turbulence ``imbalance''.
 If $\sqrt{w\pm/w_\mp}<I_{\rm max}$, then ${\cal R}=0$ and the reflection term is not applied.

Now, knowing  the dissipation of the \alf turbulence,
we are able to write the expression for ion and electron 
heating due to turbulence
\begin{equation}
Q_i = f_p \left(\Gamma_-w_- +\Gamma_+w_+\right),\quad
Q_e = (1-f_p)\left(\Gamma_-w_- +\Gamma_+w_+\right),
\end{equation}
where $f_p\approx0.6$ is a fraction of energy dissipated to ions.
 Finally, to close the system of equations, we use the following
boundary condition for the Poynting flux, $\Pi$:
\begin{equation}
\label{eq:PiToBRatio}
\frac{\Pi}{B} =\frac{\Pi_{R_\odot}}{B_{R_\odot}} = {\rm const }\approx
1.1\cdot10^6\frac{\rm W}{{\rm m}^2{\rm T}}
\end{equation}

The scaling law for the transverse correlation length:
\begin{equation}
L_\perp\sim B^{-1/2},\quad
100{\rm km}\cdot{\rm T}^{1/2}\leq L_\perp\sqrt{B}\leq
300{\rm km}\cdot{\rm T}^{1/2}
\end{equation}
\subsection{Alfv{\'e}n-Wave-Turbulence-Based Model for the Solar Atmosphere in Real Time.}
\label{sec:MHD:AWSOMR}
AWSoM has been demonstrated to be an accurate tool for modeling realistic
conditions of solar wind
\citep{Sokolov2013, Oran2013, Holst2014}.
However, in terms of computational efficiency, the model is somewhat
restrictive.
The reason for that deficiency
is the extremely fine resolution of the computational
mesh close to the solar surface;
such fine mesh is needed to resolve the dynamics of \alf wave
turbulence and ensure the numerical stability.
An alternative approach is to reformulate the mathematical problem
in the said region.
Instead of solving a computationally expensive 3-D problem
on such fine grid,
we substitute it with a multitude of much simpler 1-D problems along 
{\it threads},
that allow bringing boundary conditions up from the solar surface 
to a height defined by the assumptions below
and are the key concept of our Threaded-Field-Line-Model (TFLM).

The main assumption in the reformulated problem is that the solar magnetic
field may be considered to be potential with high accuracy
in a certain range of radii, $R_\odot<R<R_{b}$.
A thread represents a field line of such field.
A 1-D problem being introduced here, concerns a flux tube that encloses
the thread.
Reduction from 3-D to 1-D is summarized below, for more details
we refer readers to \citet{sokolov16}.
Due to the constraint on the magnetic field divergence,
$\nabla\cdot\mathbf{B}=0$, the magnetic flux remains constant
along the thread:
\begin{equation}
  \label{eq:awsomr:flux}
  B(s)\cdot A(s)={\rm const},
\end{equation}
hereafter $s$ is the distance along the field line,
$B$ is the magnitude of the magnetic field,
$A$ is the cross-section area of the flux tube in the consideration.
Conservation laws are also greatly simplified due to the fact that
in low-beta plasma, velocity is aligned with the magnetic field.
Here, assuming steady-state, conservation laws take the form:

Continuity equation:
\begin{equation}
\frac{\partial}{\partial s}\left(\frac{\rho u}{B}\right)=0\quad\Rightarrow\quad
\left(\frac{\rho u}{B}\right)={\rm const}
\end{equation}

Conservation of momentum:
\begin{equation}
\quad\frac{\partial P}{\partial s}=
-\frac{b_RGM_\odot\rho}{R^2}\quad\Rightarrow\quad
P=P_{TR}e^{\int\limits_{R_{TR}}^R
\frac{GM_\odot m_p}{2k_BT}d\frac{1}{R^\prime}},
\end{equation}
here $R_{TR}$ is the height of the transition region (TR),
$b_R$ is the radial component of $\mathbf{b}$
terms proportional to $u^2$ are neglected, 
$\mathbf{j}\times\mathbf{B}$ is omitted due to electric current vanishing
in the potential field ($\mathbf{j}\propto\nabla\times\mathbf{B}=0$)
and pressure of \alf wave turbulence is assumed to be much smaller than
the thermal pressure, $P_A\ll P$.

Conservation of energy:
\begin{eqnarray}
\frac{2N_ik_B}{B\left(\gamma-1\right)}\frac{\partial T}{\partial t} + 
\frac{2k_B\gamma}{\gamma-1}\left[\frac{N_iu}{B}\right]
\frac{\partial T}{\partial s} =\nonumber\\
\frac{\partial}{\partial s}\left(\frac{\kappa_\|}{B}
\frac{\partial T}{\partial s}\right) + 
\frac{\Gamma_-w_-+\Gamma_+w_+-N_eN_i\Lambda(T)}{B}+
\left[\frac{\rho u}{B}\right]
\frac{\partial \left(GM_\odot/R\right)}{\partial s},
\end{eqnarray}
the term $\frac{\partial T}{\partial t}$ is retained under assumption
that the electron heat conduction is a relatively slow process.
\alf wave dynamics is reformulated as well.
In Eq.~\ref{eq:w_pm} we substitute
$w_\pm=\left[\frac{\Pi}{B}\right]\sqrt{\mu_0\rho}a_\pm^2$:
\begin{equation}
\frac{\partial a_\pm^2}{\partial t}+\nabla\cdot\left(\mathbf{u}a_\pm^2\right)
\pm\left(\mathbf{V}_A\cdot\nabla\right)a_\pm^2=
\mp\mathcal{R}a_-a_+-2
\sqrt{\frac{\left[\Pi/B\right]\mu_0V_A}{\left[L_\perp\sqrt{B}\right]^2}}
a_\mp a_\pm^2
\end{equation}
The equations are additionally 
simplified since in the lower corona environment $u\ll V_A$,
i.e. waves are assumed to travel fast and quickly converge to equilibrium,
$\frac{\partial a_\pm^2}{\partial t}=0$:
\begin{equation}
\pm\left(\mathbf{b}\cdot\nabla\right)a_\pm^2=
\mp\frac{\mathcal{R}}{V_A}a_-a_+-2
\sqrt{\frac{\left[\Pi/B\right]\mu_0}{\left[L_\perp\sqrt{B}\right]^2V_A}}
a_\mp a_\pm^2
\end{equation}
Additionally, we substitute 
$d\xi=ds\sqrt{\frac{\left[\Pi/B\right]\mu_0}{\left[L_\perp\sqrt{B}\right]^2V_A}}$:
\begin{equation}
\pm\frac{da_\pm}{d\xi}=\mp\frac{ds}{d\xi}
\frac{\mathcal{R}}{2V_A}a_\mp-a_-a_+
\end{equation}

In order to close the system of equations we need to 
define the boundary conditions for TFLM.
For ``+'' wave one needs to provide value at $\xi=\xi_-$, $a_{+0}$,
and for ``-'' wave - value at $\xi=\xi_+$, $a_{-0}$.
Specifically, at the photosphere level, as the result 
from Eq.~\ref{eq:PiToBRatio} the dimensionless amplitude of the outgoing wave 
is equal to one: $b_r|_{R=R_\odot}>0:\,a_+=1\,$, $b_r|_{R=R_\odot}<0:\,a_-=1$.
Boundary conditions at the interface between TFLM and 
global corona model (GCM) are:
\bea
\left.b_r\right|_{R=R_b}>0&:&\quad
\left(\frac{u}{B}\right)_{TFLM}=\quad
\left(\frac{\mathbf{u}\cdot\mathbf{B}}{B^2}\right)_{GCM};\nonumber\\
&&\quad\left(a_-\right)_{TFLM}= \left(a_-\right)_{GCM};\quad
\left(a_+\right)_{GCM}= \left(a_+\right)_{TFLM}\nonumber\\
\left.b_r\right|_{R=R_b}<0&:&\quad
\left(\frac{u}{B}\right)_{TFLM}=-
\left(\frac{\mathbf{u}\cdot\mathbf{B}}{B^2}\right)_{GCM};\nonumber\\
&&\quad\left(a_+\right)_{TFLM}= \left(a_+\right)_{GCM};\quad
\left(a_-\right)_{GCM}= \left(a_-\right)_{TFLM}
\eea

Also one needs to sew temperature and density across the interface between
TFLM and GCM. 
We assume that the radial component of the temperature gradient is 
the dominant one, then:
\begin{equation}
\left(\frac{\partial T}{\partial R}\right)_{GCM}=
\left(\frac{\partial T}{\partial s}\right)_{TFLM}/\left|b_R\right|
\end{equation}

Boundary condition for density is controlled by sign of $u$:
\bea
{\rm for}\quad u>0&:&\quad\left(\frac{N_iu}{B}\right)_{TFLM}=
\left(N_i\right)_{TFLM}\left(\frac{u}{B}\right)_{GCM};
\nonumber\\
{\rm for}\quad u<0&:&\quad\left(\frac{N_iu}{B}\right)_{TFLM}=
\left(\frac{N_iu}{B}\right)_{GCM}.
\eea

Now we close the problem by stating conditions at the lower boundary,
i.e. at the top of TR.
Assuming steady-state, the energy conservation equation with only dominant terms
retained reads:
\begin{equation}
\frac\partial{\partial s}\left(\kappa_0T^{5/2}\frac{\partial T}{\partial s}
\right) =N_eN_i\Lambda(T),
\end{equation}
where $\kappa_\|=\kappa_0T^{5/2}$.

For a chosen width of TR along the field line,
$L_{TR}=\int\limits_{r_\odot}^{r_{TR}}ds$, 
and for a given temperature on top of the TR, $T_{TR}$, 
one can solve the heat flux and pressure from the following equations:
\bea
[N_ik_BT]&=&\frac1{L_{TR}}\int_{T_{ch}}^{T_{TR}}{\frac{\kappa_0\tilde{T}^{5/2}d\tilde{T}}
{U_{\rm heat}(\tilde{T}) }}\nonumber\\
\kappa_0 T_{TR}^{5/2}
\left(\frac{\partial T}{\partial s}\right)_{T=T_{TR}}&=&[N_ik_BT]U_{\rm heat}(T_{TR})
\eea
$\Lambda(T)$ and $U_{\rm heat}(T)=\sqrt{\frac2{k_B^2}\int^{T}_{T_{ch}}{  
\kappa_0(T^\prime)^{1/2}\Lambda(T^\prime)dT^\prime}}$ are easy to tabulate
using CHIANTI database, 
$T_{ch}\approx(1\div2)10^4K$.

Thus, TFLM is fully described as a closed mathematical problem 
that can be solved numerically.
\section{CME  Models in  Numerical Simulations}
\label{sec:MHD:CME}

As mentioned above, 
we focus in our work on gradual SEP events.
Events of this kind are characterized by a steadily increasing particle flux,
unlike impulsive events, which have an abrupt time profile \citep{reames99}.
Based on numerous observations \citep{kahler78},
it is commonly accepted that gradual, proton-rich SEP events
are associated with CMEs.
The two phenomena are linked via interplanetary shock wave, which forms
in front of a CME: the shock wave itself results from interaction of a CME
with ambient solar wind plasma and at the same time,
as shock moves outwards, it accelerates more and more particles,
hence the gradual nature of events.

Thus, properties of gradual SEP events are strongly influenced by CMEs
that trigger them.
Therefore, in order to successfully design a predictive model for gradual SEP
events, we need an accurate model to describe CMEs.
Due to the lack of 
in-situ measurements of the shock
waves and the excited turbulence in their vicinity,
numerical simulations remain the primary means of research.
A series of numerical studies employing the theory of DSA 
were performed in cases of both
idealized \citep[]{Zank2000,Rice2003,li03} and 
realistic \citep[]{Sokolov2004,Kota2005}
CME-driven shock waves.

While there are many models of CME initiation by magnetic free energy, 
these simulations are often performed in a small Cartesian box 
\citep[e.g.][]{Torok2005}, or using global models with no solar wind 
\citep[e.g.][]{Antiochos1999, Fan2004}.
So far, there have only been a few magnetically driven Sun-to-Earth 
CME simulations through a realistic interplanetary medium using 3-D MHD 
\citep[see][]{manchester04b,Manchester2005,Lugaz2007,Toth2007}.
The MHD simulation of \cite{Toth2007} was able to match the CME arrival time 
to Earth within 1.8 hours and reproduce the magnetic field magnitude 
of the event. 

In general, the purpose of a CME generator is to enhance {\it locally} the free magnetic energy of the existing {\it global} ("ambient") solution 
describing the steady state of the SC and IH magnetic field, 
${\bf B}_{\rm amb}({\bf R})$,  
by superposing an erupting configuration representing a CME's ejecta. A choice of a reasonable representation of the latter is still debatable.
 A simple but convenient  way to simulate a magnetically-driven CME is 
to superimpose  
magnetic flux-rope configuration onto the background state of the SC.
Such magnetic configuration describes an erupting magnetic filament 
filled with a plasma of excessive density.
That filament becomes an expanding flux rope (magnetic cloud) 
in the ambient solar wind while evolving and propagating outward from the Sun,
 thus allowing the simulation of the propagation to 1 AU 
of a magnetically driven CME.
In this paper, we provide a brief discussion of 
several approaches to generate CMEs with this technique.

\subsection{Magnetized cone model}
\label{sec:SSS_CM}
Observations of halo CMEs
\citep[e.g. with LASCO instrument,][]{brueckner95,plunkett98}
provided new insights into the geometry
of CMEs and its relation with other properties.
One can accurately infer the angular width and the central position
angle of a halo CME together with the plasma velocity.
For example, these observations have revealed that: 
(i) the bulk velocity tends to be radial;
(ii) the angular width, $2\Delta \theta$, tends to remain constant
as CME propagates through the corona.
These persistent features lead to the development of the 
cone model \citep{Zhao2002}.
Having only three free parameters, angular width of a CME and its initial
position on the solar surface, the model 
approximates a CME and its propagation with a cone with 
apex located at the center of the Sun.
It was later improved by \citet{Michalek2006} for arbitrary shapes.
The cone model is successfully used at 
Community Coordinated Modeling Center (CCMC) and has proved to be
an efficient tool for predicting arrival times of CMEs \citep{vrsnak14,mays15}.
Thanks to the model's accuracy and robustness,
it is used together with WSA model in CCMC's operational activities.
However, by design, the cone model lacks details about the magnetic
field carried by a CME. 
The model may be substantially enriched
as we suggest below.

In order for the simulated CME ejecta
to truly represent a magnetic cloud,
one needs to incorporate magnetic field, 
controlled by the ambient external field 
at the location where CME is added, 
into the model. 
One possible way to achieve this is to impose a spheromak, 
i.e. an equilibrium spherical MHD configuration 
(see Appendix~\ref{sec:Spheromak}),
around the central point of the cloud, ${\bf R}_{\rm c}$. 
Spheromak's magnetic field is:
\begin{equation}
\label{eq:SpheromakBeta}
{\bf B}_{\rm sk}({\bf r})=\left[\frac{j_1({\alpha_0}{r})}{\alpha_0{r}}-\beta_0\right]\left(2{\bf B}_0+\sigma_h \alpha_0[{\bf B}_0\times{\bf r}]\right)
+j_{2}(\alpha_0{r})\frac{[{\bf r}\times[{\bf r}\times{\bf B}_0]]}{r^2},
\end{equation}
where $j_1$ and $j_2$ are spherical Bessel functions.
Herewith, the vector ${\bf B}_0$ is introduced with the magnitude equal 
to $B_0$ directed along spheromak's axis of symetry,
$\sigma_h=\pm1$ is the sign of helicity (we assume $\alpha_0={\rm const} >0$),
$\beta_0={\rm const}$ is the charactersitic value of plasma beta. 
The coordinate vector, ${\bf r}$, originates at the center of configuration, 
$\mathbf{R}_{\rm c}$. 
We assume no currents outside a spherical magnetic surface 
$\|{\bf R}-{\bf R}_{\rm c}\|=r_0$, which thus bounds the configuration. 
The radial and toroidal components of the magnetic field turn 
to zero at the surface, thus $j_1(\alpha_0r_0)=\beta_0\alpha_0r_0$. 
For a given $\beta_0$ this equation relates the configuration size, $r_0$, to the extent of magnetic field twisting, $\alpha_0$, 
needed to close the configuration within this size.

One also needs to account for the field, which the currents 
{\it inside} spheromak produce {\it outside} the boundary, 
$\|{\bf R}-{\bf R}_{\rm c}\|=r_0$. 
The calculation of
the magnetic moment \citep[see definition in ][]{jackson1999}, $\mathfrak{m}$,
of the spheromak configuration gives:
\bea
\mathfrak{m}=\frac12\int\limits_{\|{\bf r}\|\le r_0}{d^3r[{\bf r}\times{\bf j}]}=\frac1{2\mu_0}\int\limits_{\|{\bf r}\|\le r_0}{d^3r[{\bf r}\times[\nabla\times{\bf B}_{\rm sk}]]}=\frac{4\pi r_0^3}{3\mu_0}j_2({\alpha_0r_0}){\bf B}_0
\eea
The final expression for $\mathfrak{m}$
is obtained via reducing the volume integral to 
the integral over the spheromak's surface,
at which \mbox{${\bf B}_{\rm sk}|_{r=r_0}=j_2(\alpha_0r_0)\frac{[{\bf r}\times[{\bf r}\times{\bf B}]]}{r^2}$}. 
The field of magnetic dipole, $\mathfrak{m}$, 
which we admit as the spheromak's field outside the boundary, equals: 
\bea
{\bf B}_{{\rm sk,}r>r_0}({\bf r})=\frac{\mu_0}{4\pi r^3}\left\{\frac{3\left({\bf r}\cdot\mathfrak{m}\right){\bf r}}{r^2}-\mathfrak{m}\right\}=j_2({\alpha_0r_0})\frac{r_0^3}{r^3}\left\{\frac{\left({\bf r}\cdot{\bf B}_0\right){\bf r}}{r^2}-\frac{{\bf  B}_0}3\right\}
\eea
Now, we provide the full expression for a spheromak superposed onto
 ambient field, $B_{\rm amb}(\mathbf{r})$ :
\bea\label{eq:ConeModel}
{\bf B} ({\bf R}) = \left\{\begin{array}{c}{\bf B}_{\rm amb} ( {\bf R} )+{\bf B}_{{\rm sk,}r>r_0}({\bf R}-{\bf R}_{\rm c}),\quad\|{\bf R}-{\bf R}_{\rm c}\|\ge r_0 \\ \left[{\bf B}_{\rm amb}( {\bf R})+\frac23j_2({\alpha_0r_0}){\bf B}_0\right]+{\bf B}_{\rm sk}({\bf R}-{\bf R}_{\rm c}),\quad\|{\bf R}-{\bf R}_{\rm c}\|\le r_0\end{array}\right.
\eea
where the uniform field, $\frac23j_2({\alpha_0r_0}){\bf B}_0$ 
is added to the spheromak field for two reasons. 
First, this preserves the field continuity at 
$\|{\bf R}-{\bf R}_{\rm c}\|=r_0$, 
i.e. 
from both side of the boundary the field equals 
$j_2({\alpha_0r_0}) \left\{\frac{\left({\bf r}\cdot{\bf B}_0\right){\bf r}}{r^2}-\frac{{\bf  B}_0}3\right\} $. 
Second, certain aspects of CME ejecta's interaction with ambient plasma dictate
this correction. 
Indeed, if an ejecta represents a {\it magnetic cloud}, its {\it frozen in} 
magnetic field effectively {\it replaces} the pre-existing field, 
$\mathbf{B}_{\rm amb}$, 
at any location, $\mathbf{R}_{\rm cloud}$, it passes. 
Therefore,the cloud's internal field, which we assume to be the 
superposition of the ambient field with the field of the spheromak
centered at $\mathbf{R}_{\rm cloud}$ 
\mbox{(i.e. $\mathbf{R}_{\rm cloud}\equiv\mathbf{R}_{\rm c}$)},
must be corrected by the negative of this pre-existing field.
This reasoning demands the expression in the square brackets in
Eq.~\ref{eq:ConeModel} be exactly zero at $\mathbf{R}_{\rm c}$, 
i.e. $\mathbf{B}_0$ and $\mathbf{B}_{\rm amb}$ must be related as:
\bea\label{eq:FieldChoice}
{\bf B}_0=-\frac3{2j_2(\alpha_0r_0)}{\bf B}_{\rm amb}({\bf R}_{\rm c})
\eea
which ensures both the continuity of the field, Eq.~\ref{eq:ConeModel}, 
and the proximity of the internal field 
(equality, if the ambient field is uniform), 
$\left[{\bf B}_{\rm amb}( {\bf R})-{\bf B}_{\rm amb}({\bf R}_{\rm c})\right]+{\bf B}_{\rm sk}({\bf R}-{\bf R}_{\rm c})$, 
to the equilibrium state 
${\bf B}_{\rm sk}({\bf R}-{\bf R}_{\rm c})$. 
Should the field of the superimposed configuration 
not match the ambient field {\it in direction}, the non-zero torque, 
$\left[\mathfrak{m}\times{\bf B}_{\rm amb}({\bf R}_{\rm c})\right]$ acting on the magnetic moment, $\mathfrak{m}$, in the field ${\bf B}_{\rm amb}({\rm R}_{\rm c})$, 
would tend to align the configuration axis with the external field.  
Should the configuration field be stronger/weaker than that governed by 
Eq.~\ref{eq:FieldChoice}, the ambient field would be too weak/strong 
to balance the hoop force in the spheromak configuration, 
so that the latter would tend to expand/shrink. 
The field in the configuration determined by 
Eqs.~\ref{eq:ConeModel},~\ref{eq:FieldChoice} is oppositely directed and 
somewhat stronger than the ambient field. 
For comparison, the field in the center of configuration equals: 
${\bf B}_{\rm sk}({\bf R}_{\rm c})=2(\frac13-\beta_0){\bf B}_0=-\frac{1-3\beta_0}{j_2(\alpha_0r_0)}\mathbf{B}_{\rm amb}(\mathbf{R}_{\rm c})$.
Magnetic geometry of 
the described configuration provides a natural explanation 
of the geomagnetic activity caused by CMEs. 
Indeed, if the configuration described above passes the Earth location, 
the local magnetic field may consequently change from 
${\bf B}_{\rm amb}({\bf R}_{\rm c})$ to ${\bf B}_{\rm sk}({\bf R}_{\rm c})$ and back, 
so that all components of the interplanetary magnetic field change sign 
and increase in absolute value by a factor of 
$\frac{(1-3\beta_0)}{j_2(\alpha_0r_0)}\approx(4\div5)$. 
This is a classical scenario for the magnetospheric storm.

Disregarding the solar gravitational pull and assuming uniform ambient field,
the magnetic configuration described above is in a force equilibrium.
As demonstrated by \citet{low82}, 
once some special distribution of plasma velocity is imposed
onto an equilibrium magnetic structure
with adiabatic index $\gamma{=}4/3$ , this structure
starts to evolve {\it self-similarly},
i.e. with the only change in its geometry being the radial motion
and uniform expansion
\citep[see][about self-similar solutions]{sedov59,zeldovich67}.
Specifically, we need to assume, 
and implement in the numerical simulations, 
the radially diverging initial motion with the radial velocity 
at each point being proportional to the heliocentric distance:
$$\mathbf{u}|_{t=0}= U_{\rm CME}\frac{\mathbf{R}}{\|\mathbf{R}_{\rm c}\|},$$
where the CME speed, $U_{\rm CME}$, may be found from observations.
In application to the magnetized cone model 
this means that superimposing a spheromak with such velocity profile
onto a barometric atmosphere would be consistent
with basic principles of the cone model of \citet{Zhao2002}:
(i) bulk velocity of the resulting magnetic cloud is radial, and 
(ii) shape of the cloud, due to self-similarity, remains constant.
As noted above, we neglected the gravity, i.e.
the moving magnetic cloud isn't in the perfect force equilibrium.
Therefore, exact self-similarity can't be achieved, 
rather it is approached when relative contribution of gravitational force tending to decelerate the cloud
is small. Another force tending to decelerate the magnetic cloud is the drag force, which opposes to the faster CME motion through the slower moving ambient. On the other hand, in non-uniform ambient magnetic field anti-parallel to $\mathbf{B}_0$, the force acting on the magnetic dipole, repels it out of the active region, thus, accelerates its radial motion. These counteracting forces may partially balance each other, thus resulting in almost steady radial motion as assumed by the cone model.  

\subsection{Stretched Spheromak Configuration by Gibson-Low}
\label{sec:SS_GL} 
 The family, (\ref{eq:SpheromakBeta}), of 
equilibrium configurations may be extended with the use of {\it coordinate transformation} suggested by \citet{gibson98}.
The arising pressure imbalance perfectly compensates
the gravitational force acting on the spheromak's plasma.
The new equilibrium configuration in the heliocentric coordinates, 
$\mathbf{R}$, 
with the magnetic field, $\mathbf{B}(\mathbf{R})$, and pressure distribution, 
$P(\mathbf{R})$, may be described in terms of the spheromak solution, (\ref{eq:SpheromakBeta}), of 
the Grad-Shafranov equations (see Appendix~\ref{sec:Spheromak}),
${\bf B}^\prime({\bf R}^\prime)=\mathbf{B}_{\rm sk}(\mathbf{R}^\prime-\mathbf{R}_{\rm c})$ and 
$P^\prime({\bf R}^\prime)=P_{\rm sk}(\mathbf{R}^\prime-\mathbf{R}_{\rm c})$.
For each point, $\mathbf{R}$, 
we take the values of these functions in the point, 
$\mathbf{R}^\prime(\mathbf{R})=\left(1+\frac{a}R\right)\mathbf{R}$,  
$R^\prime=R+a$,  which is radial coordinate stretching, 
an arbitrary constant $a$ being the distance of stretching.
When the stretching transformation is applied, it displaces the magnetic 
configuration toward the heliocenter and gives it a teardrop-like shape.
The magnetic field  vector in the course of stretching should be scaled 
in addition to the coordinate transformation: 
\be
\label{eq:GL:B}
\mathbf{B}(\mathbf{R})=
\frac{R^\prime}R\left(\mathbb{I}+\frac{a}{R}\mathbf{e}_R\mathbf{e}_R\right)\cdot\mathbf{B}^\prime(\mathbf{R}^\prime)
\ee
where $\mathbf{e}_R=\mathbf{R}/R$ and $\mathbb{I}$ is the identity matrix. 
The radial field component, 
$B^\prime_{R}=\left(\mathbf{B}^\prime\cdot\mathbf{e}_{R}\right)$, 
is thus multiplied by $\left(\frac{R^\prime}R\right)^2$, 
all the other by $\frac{R^\prime}R$. 
Thus transformed magnetic field is divergence-free. 
The plasma pressure of the stretched magnetic configuration is defined as:
\begin{equation}
\label{eq:GL:P}
P(\mathbf{R}) = \left(\frac{{R^\prime}}{R}\right)^2\left(P^\prime
-\frac{a}R\left(2+\frac{a}R\right)
\frac{B^{\prime\,2}_R}{2\mu_0}\right)
\end{equation} 
One can verify an equilibrium condition  
for the transformed magnetic configuration. 
The spatial derivatives of $\mathbf{B}^\prime(\mathbf{R}^\prime)$ and 
$P^\prime(\mathbf{R}^\prime)$ are transformed as follows: 
$\nabla=\left[\left(1+\frac{a}R\right)\mathbb{I}-\frac{a}R\mathbf{e}_R\mathbf{e}_R\right]\cdot\nabla^\prime$. 
Using the equilibrium condition for a non-stretched configuration, 
$\frac1{\mu_0}[[\nabla^\prime\times\mathbf{B}^\prime]\times\mathbf{B}^\prime]-\nabla^\prime P^\prime=0$, 
the left hand side of Eq.~\ref{eq:GL:start} may be reduced 
to the following form: \mbox{$\frac1{\mu_0}\left[
\left[
\nabla\times\mathbf{B}
\right]\times\mathbf{B}\right]
-\nabla P=F_R\mathbf{e}_R$}, 
where the radial force arising from extra tension of the stretched magnetic field is:
\begin{equation}
\label{eq:GL:RadialForce}
F_R = \frac{{a}R^{\prime\,2}}{R^3}\left[
\left({2}+\frac{a}R\right)
\left(
\frac{B^{\prime\,2}}{\mu_0R^\prime}+\left(\mathbf{e}_R\cdot\nabla^\prime\right)
\left(P^\prime +\frac{B^{\prime\,2}}{2\mu_0}\right)
\right)
+\frac{2P^\prime}{R^\prime} - 
\left(3+2\frac{a}R\right)
\frac{B^{\prime\,2}_R}{\mu_0R}
\right]
\end{equation}
Now, one can consider the stretched magnetic configuration described 
by Eqs.~\ref{eq:GL:B}, ~\ref{eq:GL:P} once superposed with some background 
barometric distribution of pressure, $P_{\rm bar}(\mathbf{R})$, and density,  
$\rho_{\rm bar}(\mathbf{R})$, 
which satisfy the hydrostatic equilibrium condition, 
$-\nabla P_{\rm bar}+\rho_{\rm bar}\mathbf{g}=0$, 
$\mathbf{g}=-GM_\odot\mathbf{e}_R/R^2$. 
The superposed distribution satisfies the equilibrium condition 
accounting for gravity:
\be
\frac1{\mu_0}\left[
\left[
\nabla\times\mathbf{B}
\right]\times\mathbf{B}\right]
-\nabla \left(P+P_{\rm bar}\right)+\left(\rho+\rho_{\rm bar}\right)\mathbf{g}=0
\ee
if the density variation due to the effect of stressed field is
\be
\label{eq:GL:Rho}
\rho = \frac{F_R}{g(R)}
\ee

\begin{figure}
\centering
\includegraphics[width={3in},height={2.5in}]{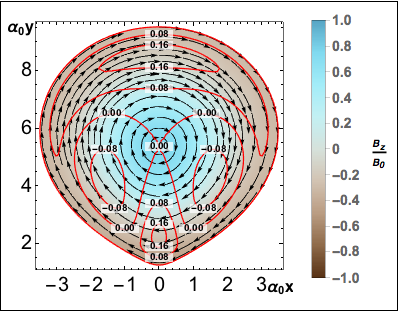}
\includegraphics[width={3in},height={2.5in}]{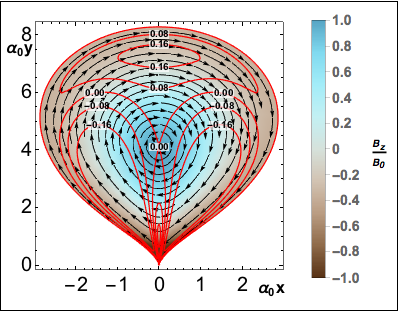}
\caption{
\small{
Equatorial plane of the stretched flux rope for $\beta_0{=}0.02$ 
(cf.~Fig.~\ref{Fig:Spheromak}). 
The original flux rope is shifted by distance $1.6r_0$ 
along a direction in the equatorial plane and then stretched towards 
the heliocenter by distance $0.3r_0$ ({\bf left}) 
and $0.6r_0$ ({\bf right}).
Magnetic field direction 
is marked with arrows, 
off-plane component of the magnetic field is normalized per $B_0$ 
(see Eq.~\ref{eq:SpheromakBeta}) and shown by color.  
Local values of  plasma parameter 
$\beta({\bf r})=\mu_0P({\bf r})/B^2({\bf r})$ 
are shown with red curves corresponding to levels 
$\beta=0.04,0.08,0.12,0.16$ as marked explicitly.
}
}
\label{Fig:GL}
\end{figure}
As a result of the transformation, the spherical configuration is stretched
towards the heliocenter as shown in Fig.~\ref{Fig:GL}.
When the solution represented by 
Eq.~\ref{eq:GL:B},~\ref{eq:GL:P},~\ref{eq:GL:Rho} (the GL flux rope)
is superimposed onto the existing corona, 
the sharper end of the teardrop shape is submerged
below the solar surface. 
In the wider top part of the configuration ("balloon") the density variation 
in Eq.~\ref{eq:GL:Rho} is {\it negative}, that is the resulting density 
is lower than that of the ambient barometric background. 
As the result, the Archimedes force acting on this part pulls 
the whole configuration outward the Sun. 
The cavity with the reduced density is often observed in the CME images 
from the LASCO coronagraphs. 
Then, in the narrower bottom part of the configuration ("basket") 
the excessive {\it positive} density simulates the dense ejecta, 
which is pulled outward the Sun by the radial tension 
in the stretched magnetic configuration. 
Finally, the tip of the configuration with the magnetic field lines 
both ingoing and outgoing the solar surface in anchored to the negative 
and positive magnetic spots of a bipolar AR, 
considered as the source of the CME. 
Depending on the reconnection rate, the configuration can either keep 
being magnetically connected to the AR, or it may disconnect and close 
and then propagate toward 1 AU as the magnetic cloud. 

The time evolution of GL flux rope
is self-similar \citep[provided $\gamma=4/3$, see][]{low82}.
Additionally, this result may be generalized:
adjusting the density profile in Eq.~\ref{eq:GL:Rho}
for effective gravity $g(R)+\alpha R$
would result in accelerated/decelerated
propagation of a CME.
A literal requirement for self-similarity of GL flux rope can hardly be fulfilled in realistic corona. Indeed,
in order for the configuration
to remain in force-equilibrium 
(or to keep the specific shape of the force imbalance to maintain acceleration)
 and therefore propagate in
a self-similar fashion,  
a specific and unrealistic distribution of the external pressure is needed.
Additionally, since the Ampere's force is non-linear in 
magnetic field, superimposing GL flux-rope adds a new  effect of the background magnetic field onto the flux-rope's currents,
which contributes even more to the force imbalance.
The significance of these effects hasn't been thoroughly studied,
however, CME propagation has been shown to be approximately self-similar
\citep[e.g.][]{manchester04a,manchester04b}.

The GL flux rope model has been used for CME initiation in, for example,
\citet{manchester04a,manchester04b,manchester06,lugaz05,Jin:2017a,Jin:2017b}. 
The recent developments allowed significant simplification of
the process of triggering CMEs using GL model.
The product of the effort is the 
Eruptive Event Generator based on Gibson-Low magnetic configuration
(EEGGL) \citep{Jin:2017b},
which is discussed in details in \cite{borovikov17}.
\subsection{Thin Flux Rope by Titov-Demoulin}
\label{SSS_TD}
The approach of TD also stems from consideration of the magnetic field's
topology.
A pre-eruptive configuration of the field is reconstructed with 3 different
components,
$\mathbf{B}_I$, $\mathbf{B}_q$, $\mathbf{B}_\theta$.
$\mathbf{B}_I$ is created by a uniform ring current flowing 
in the emerging flux rope (later the model has been modified in \citet{titov14}
to include a non-uniform current profile, TDm hereafter),
$\mathbf{B}_q$ is the magnetic field of two equal imaginary 
magnetic charges of opposite signs embedded below the solar surface and, 
finally,
$\mathbf{B}_\theta$ is produced by a constant line current flowing through 
the said charges.

The TD flux rope model has been used in a number of studies
\citep[][e.g.]{Roussev2003a,manchester08}, as well as its modified version,
TDm \citep{Linker2016}.
Specific examples of CME simulations using the AWSoM model 
for the SC and IH with a superimposed TD magnetic configuration include 
\citet{Manchester2012} and \citet{Jin2013}.

{
\section{Interface between MHD and Kinetic Models}
\label{sec:kinetic}
\subsection{Transport equation}
\label{sec:kinetic:transport}
The transport of energetic particles through the inter-planetary space
by itself is
an important problem in space science.
It was studied since the discovery of the Galactic Cosmic Rays (GCR), 
the energetic particles originating
from beyond the Solar system.
A comprehensive summary of the problem can be found in the review by
\citet{Parker1965}.
Although results in the said review are 
obtained in a different context,
some can readily be applied for the SEP transport.

The distribution of SEPs is far from Maxwellian, therefore, 
they should be characterized by a (canonical) distribution function 
$F({\bf R},{\bf p},t)$ of coordinates, ${\bf R}$, and momentum, ${\bf p}$, 
as well as time, $t$, such that the number of particles, $dN$, 
within the elementary volume,  $d^3{\bf R}$, 
is given by the following integral: 
$dN=d^3{\bf R}\int{d^3{\bf p}\,F({\bf R},{\bf p},t)}$. 
In a magnetized plasma, 
it is convenient to deal with the distribution function at the given point, 
${\bf R}$, in the co-moving frame of reference, 
which moves with the local speed of interplanetary plasma, 
${\bf u}({\bf R},t)$, on introducing spherical coordinates, 
$(p=|{\bf p}|,\mu={\bf b}\cdot{\bf p}/p,\varphi)$ 
in the momentum space with its polar axis aligned with the direction 
of the magnetic field, ${\bf b}$, 
herewith $\mu$ being the cosine of pitch-angle.
The normalization integral in these new variables becomes: 
$dN=d^3{\bf R}\int_0^\infty{p^2dp\int_{-1}^1{d\mu\int_0^{2\pi}{d\varphi F({\bf R}, p,\mu,\varphi,t)}}}$. 
Using this canonical distribution function, 
one can also define a gyrotropic distribution function, 
$f({\bf R}, p,\mu,t)=\frac1{2\pi}\int_0^{2\pi}{d\varphi F({\bf R}, p,\mu,\varphi,t)}$. 
This function is designed to describe the particle motion averaged over 
the phase of its gyration about the magnetic field.  
The isotropic (omnidirectional) distribution function, 
$f_0({\bf R}, p,t)=\frac1{2}\int_{-1}^1{d\mu f({\bf R}, p,\mu,t)}$ 
is averaged over the pitch angle too. 
The normalization integrals are: 
$dN=2\pi d^3{\bf R}\int_0^\infty{p^2dp\int_{-1}^1{d\mu f({\bf R}, p,t)}}=4\pi d^3{\bf R}\int_0^\infty{p^2dp f_0({\bf R}, p,t)}$
 
The commonly used kinetic equation for the isotropic part of 
the distribution function has been introduced in \citet{Parker1965}:
\begin{equation}
\label{eq:parker}
\frac{\partial}{\partial t}f_0\left(\mathbf{R},p,t\right) + 
\left(\mathbf{u}\cdot\nabla\right)f_0\left(\mathbf{R},p,t\right) -
\frac{1}{3}\left(\nabla\cdot\mathbf{u}\right)
\frac{\partial}{\partial\ln p}f_0\left(\mathbf{R},p,t\right) = 
\nabla\cdot\left(\kappa\cdot\nabla f_0\left(\mathbf{R},p,t\right)\right) + S,
\end{equation}
where $\varkappa=D_{xx}\mathbf{b}\mathbf{b}$ 
is the tensor of parallel (spatial) diffusion along the magnetic field, 
$S$ is the source term.
In this approximation,  the cross-field diffusion of particles is neglected.

Eq.~\ref{eq:parker} captures the effect of interplanetary plasma and IMF
on the SEP transport and acceleration. 
The term proportional to the divergence of $\mathbf{u}$ is the 
adiabatic cooling, for $\left(\nabla\cdot\mathbf{u}\right)>0$,  
or (the first order Fermi) acceleration in compression or shock waves.
According to estimates by \cite{Parker1965}, 
during quiet time the adiabatic scaling of particles' energy from 
their origin to 1~AU is 
$\propto\left(\rho_{1AU}/\rho_\odot\right)^{\left(n/3\right)}$,
where $n{=}2$ for non-relativistic and $n{=}1$ for relativistic particles.

Small scale irregularities also have a significant impact on particle 
propagation.
Their scale is ${\sim}10^5{\div}10^7$~km, which is comparable
with gyroradii of SEP but very small compared to 1~AU.
Particles scatter on these irregularities, and on the large scale the particle motion can be described as diffusion,
the first term on the right in Eq.~\ref{eq:parker}. 
Based on Eq.~\ref{eq:parker}, \cite{Krymsky1977,Axford1977,Bell1978a,Bell1978b,Blandford1978} 
proposed  the (DSA)  mechanism
to explain the observed power-law spectra of GCRs.

In the present paper we limit our consideration with the case of the Parker equation~\ref{eq:parker} as the model to describe the SEP acceleration and transport. More realistic and accurate models accountic for the pitch-angle dependence for the distribution function are delegated to the companion paper.

\subsection{Lagrangian coordinates and Field Line Advection Model}
\label{sec:Lagrangian}
We adopt Eq.~\ref{eq:parker} as mathematical approach to the problem
of SEP transport. However, this consideration is computationally challenging:
a fully 3-D propagation of particles requires significant resources.
This can be avoided by observing that Eq.~\ref{eq:parker} assumes that the particle motion in physical
space consists of the particle guiding center's displacement along the interplanetary magnetic field (IMF) and advection with plasma
into which the IMF is frozen. 
This property allows us to describe the particle propagation in the Lagrangian coordinates. 
The benefits of this approach is the reduction of a complex 3-D problem to a multitude of much simpler 1-D problems along magnetic field lines, with no loss of generality.

At the early age of the mechanics of continuous media there were 
two competing approaches to a mathematical description of the motion of 
fluids. 
In Eulerian coordinates, ${\bf R}, t$, 
the distribution of the fluid parameters 
(density, velocity, temperature, pressure, etc) 
at each instant of time, $t$, is provided as a function of coordinates, 
${\bf R}$, in some coordinate frame. 
No need to emphasize that the any given point, ${\bf R}$ is immovable, 
while the fluid passes this point with the local flow velocity 
${\bf u}({\bf R},t)$, 
so that at each time instant the fluid element at this point differs 
from that present at this point a while ago. 
In contrast with this approach, the Lagrangian coordinates, ${\bf R}_L$, 
stay with the given fluid element rather than with 
the given position in space. 
While the fluid moves, each moving fluid element keeps unchanged 
the value of the Lagrangian coordinates, ${\bf R}_L$, 
while its spatial location, ${\bf R}\left({\bf R}_L,t\right)$, 
changes in time in accordance with the definition of the local fluid velocity:
\begin{equation}
\label{eq:DxDt}
\frac{D{\bf R}({\bf R}_L,t)}{Dt}={\bf u}({\bf R},t)
\end{equation}
Here, the partial time derivative at constant Lagrangian coordinates, ${\bf R}_L$ is denoted as $\frac{D}{Dt}$, while the usual notation, $\frac\partial{\partial t}$, 
denotes the partial time derivative at constant Eulerian coordinates, ${\bf R}$. 
As usually, we choose the Lagrangian coordinates for a given fluid element equal to the Eulerian coordinates of this element at the initial time instant, ${\bf R}_L={\bf R}|_{t=0}$. For numerical simulations, with any choice of the grid in Lagrangian coordinates, $\left({\bf R}_{ijk}\right)_L=  \left({\bf R}_{ijk}\right)|_{t=0}$, one can numerically solve the multitude of ordinary differential equations, Eq.~\ref{eq:DxDt}, to trace the spatial location for all Lagrangian grid points in the evolving fluid velocity field, ${\bf u}({\bf R},t)$, as long as the latter is known. 

An example of application of Lagrangian coordinates to the Parker equation,
 Eq.~\ref{eq:parker},
is FLAMPA \citep{Sokolov2004}.
\subsection{M-FLAMPA}
\label{sec:MFLAMPA}
Geometry of magnetic field lines may become very complex and
they can form intricate patterns as they evolve in time.
By pushing and twisting field lines, extreme events, such as CMEs and
associated interplanetary shocks, can make the field line topology
even more complex.
This makes forecasting the regions affected by  SEP events
a challenging problem.
To address this challenge one needs to design a computational technique
that naturally and efficiently describes this ever evolving geometry.
The Multiple-Field-Line-Advection Model for Particle Acceleration 
(M-FLAMPA)
 code was designed to solve this problem.
M-FLAMPA allows us to solve the kinetic equation for SEPs
along a multitude
of interplanetary magnetic field lines originating from the Sun,
using time-dependent magnetic field and plasma parameters obtained
from the MHD simulation.
The model is  a high-performance extension of the original FLAMPA code
\citep{Sokolov2004},
which simulates SEP distribution along a single field line.
M-FLAMPA is a major improvement that takes full advantage of
modern supercomputers.

M-FLAMPA solves for gyrotropic
SEP distribution function $f(\mathbf{x}, p, t)$,
where $p$ is the magnitude of the relativistic momentum of energetic particles.
The code takes advantage of the fact that particles stay
on the same magnetic field line and, therefore, the distribution function
may be treated as a function of the distance along the filed line, $s$,
rather than a 3-D vector $\mathbf{x}$.
Also, coefficients in the governing equations depend only on background
plasma parameters and their Lagrangian derivatives
(see Section~\ref{sec:Lagrangian}).
This important property reduces the problem of particle acceleration
in 3-D magnetic field into a set of independent 1-D problems
on continuously evolving Lagrangian grids.
In other words, each field line in the model is treated separately from others, which results
in a perfectly parallel algorithm.
We note that the same computational technology is applied
to the transport equations for the \alf wave amplitudes \citep{Sokolov2009}.

M-FLAMPA is directly coupled with SC and IH MHD models via 
an advanced coupling algorithm within the SWMF.
This technique seamlessly connects field lines between
the two distinct computational domains, 
where lines are extracted based on a concurrently updated solution
of solar wind parameters.
The line extracting procedure is augmented with a new 
interpolation algorithm \citep{Borovikov2015} that eliminates
spurious distortions near grid resolution interfaces
that routinely occur in large scale MHD simulations.
The underlying algorithmic innovations ensure that MFLAMPA can combine
the accuracy of realistic MHD simulations with high computational efficiency.
Thus, the new technology is well suited for modeling and predicting SEP impacts during extreme solar events.

The integrated model traces magnetic field lines from the MHD models
to find the area that is covered by field lines originating
from a given area of the solar surface, such as an active region.
As described above, 
each field line is represented by a Lagrangian grid that
advects with the background plasma in a time dependent manner.
The relevant data at the location of the grid points is transferred to
MFLAMPA, which in turn calculates the evolution of the energetic particle
population by solving the governing kinetic equations.

}
{
\section{Proof-of-concept results from the MHD+SEP coupled model}
\label{sec:Results}
\subsection{Simulation of SEP event of January 23, 2012}
\label{sec:Jan23}
As a demonstration of our predictive framework's capabilities
we provide simulation results for the SEP event
associated with the CME observed of January 23, 2012.
Various aspects of the event have been studied in the literature,
e.g. \citet{tyssoy13,joshi13}.
The simulation is performed as follows:
(1) use magnetogram for late January 2012 (Carrington rotation 2119\footnote{Available at \url{https://gong.nso.edu/data/magmap/crmap.html}})
to find a pre-eruptive, steady state solution for SC and IH using;
(2) initiate a CME with parameters computed by EEGGL tool\footnote{Available at \url{https://ccmc.gsfc.nasa.gov/eeggl/}} for anticipated CME speed, e.g. as measured by StereoCAT tool\footnote{Available at \url{https://ccmc.gsfc.nasa.gov/stereocat/}} 
or found DONKI Space Weather activity archive\footnote{Available at \url{https://kauai.ccmc.gsfc.nasa.gov/DONKI/}};
(3) run MHD and particle models concurrently, where the former provides
background solar wind parameters for the latter.

\begin{figure}[!t]
  \centering
  \includegraphics[width=0.3\linewidth]{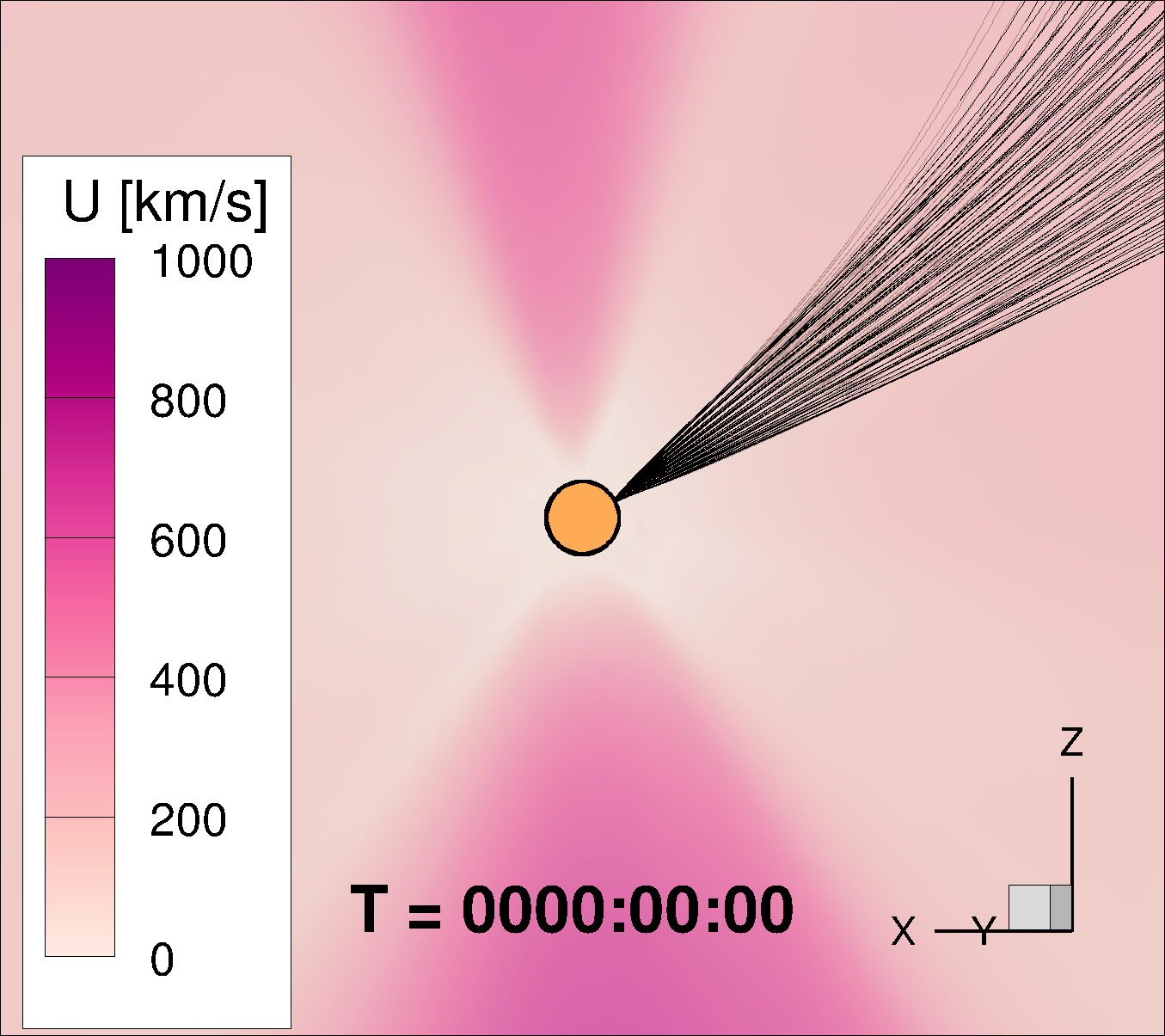}
  \includegraphics[width=0.3\linewidth]{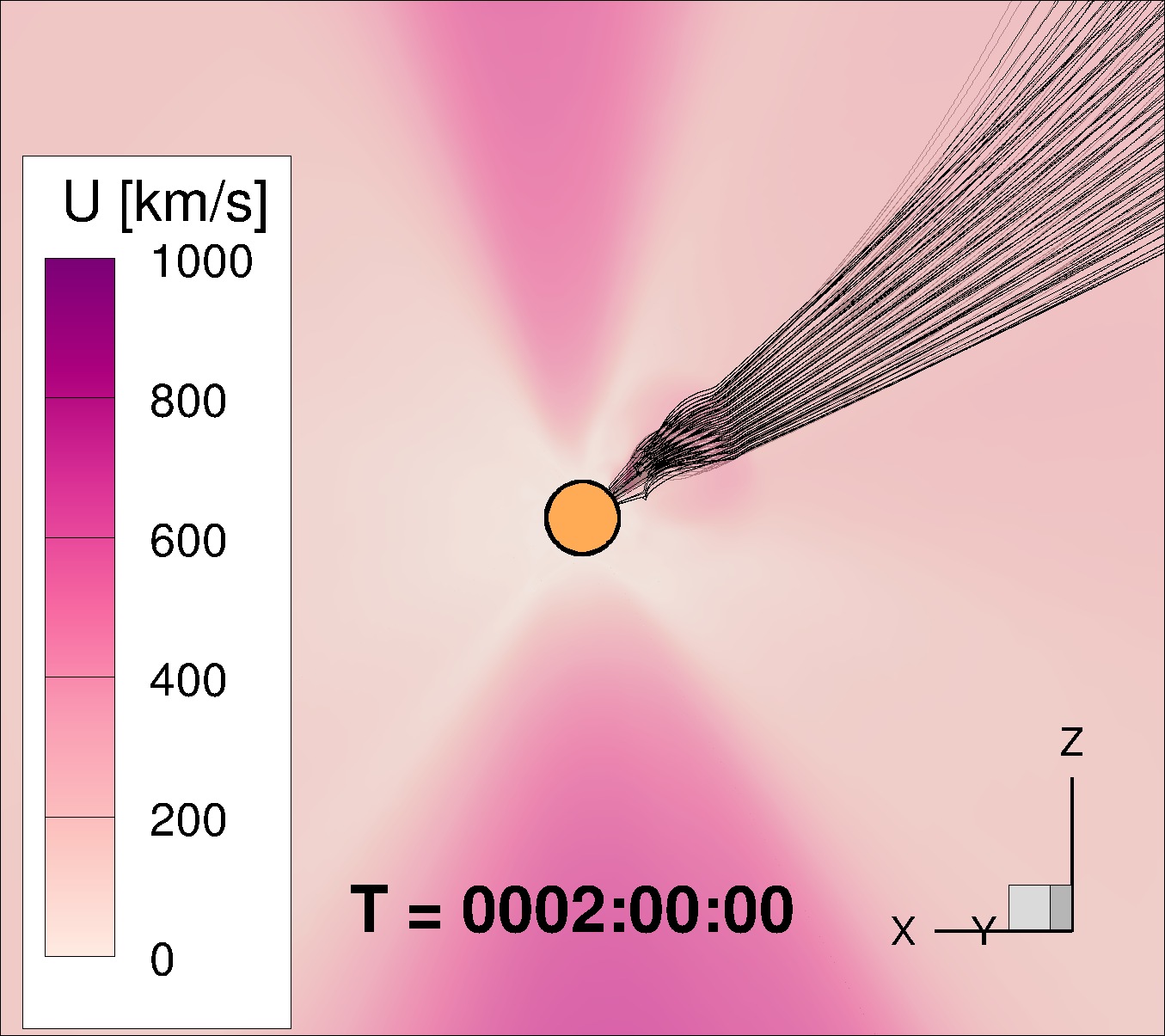}
  \includegraphics[width=0.3\linewidth]{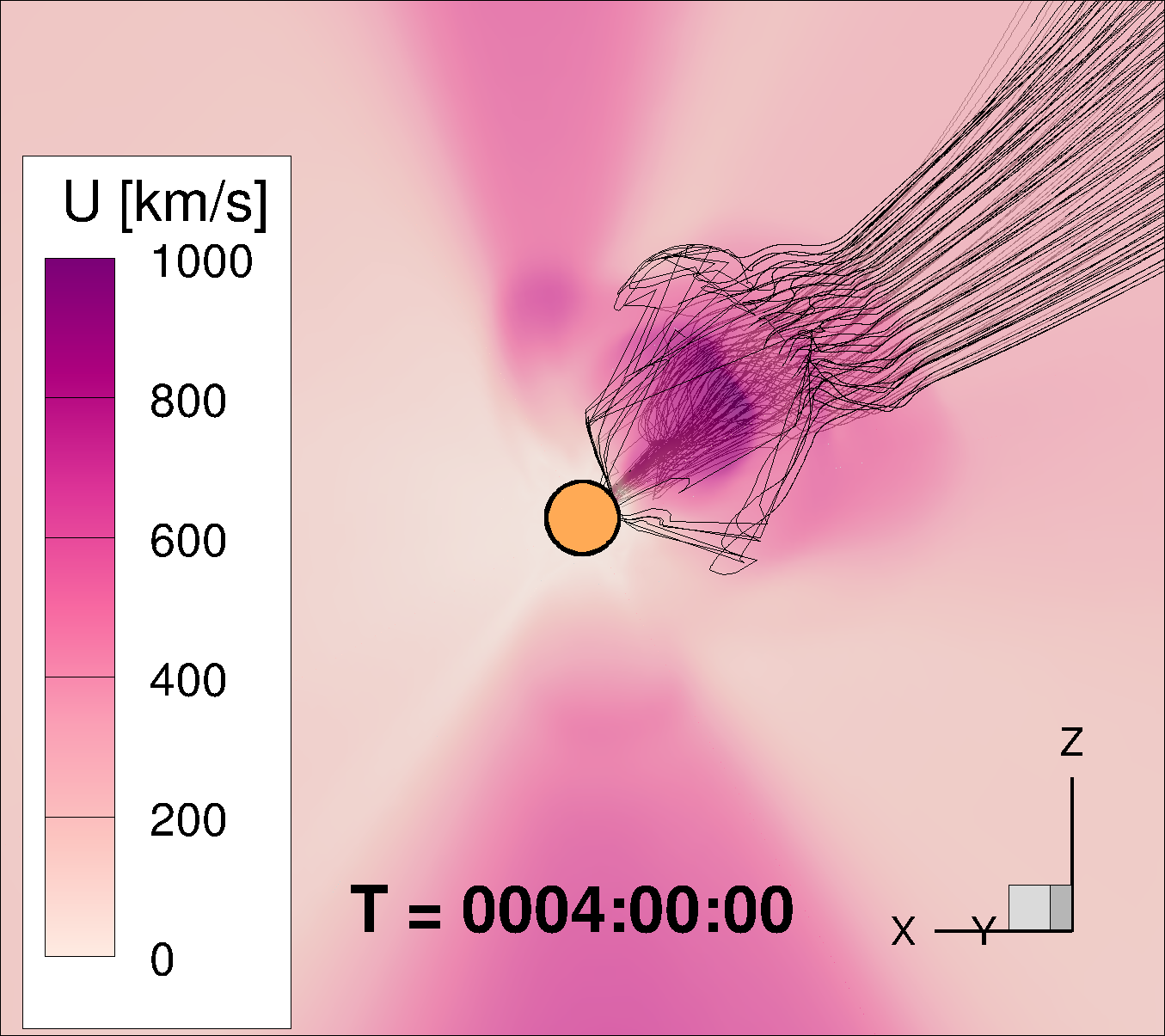}\\
  \includegraphics[width=0.3\linewidth]{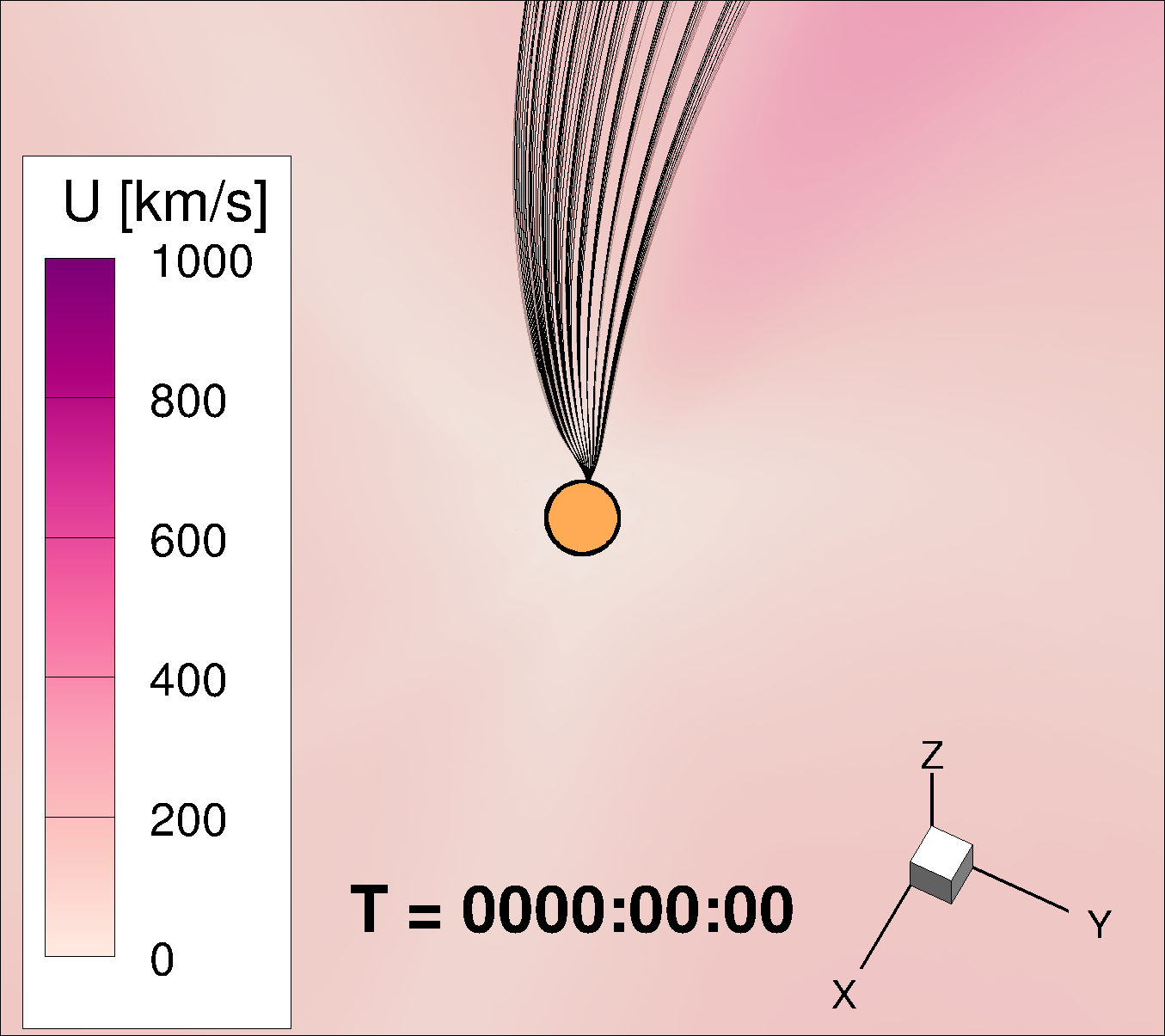}
  \includegraphics[width=0.3\linewidth]{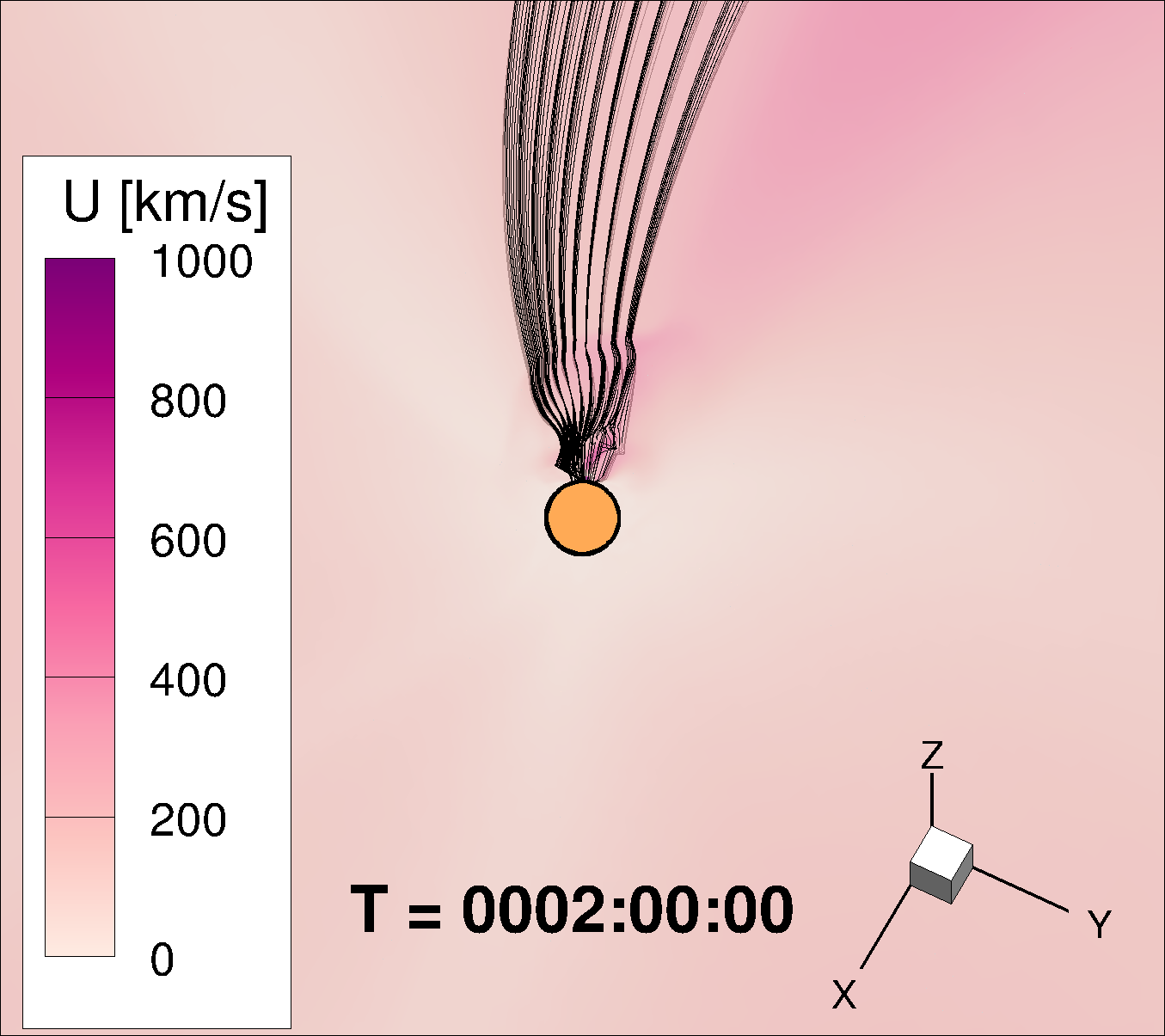}
  \includegraphics[width=0.3\linewidth]{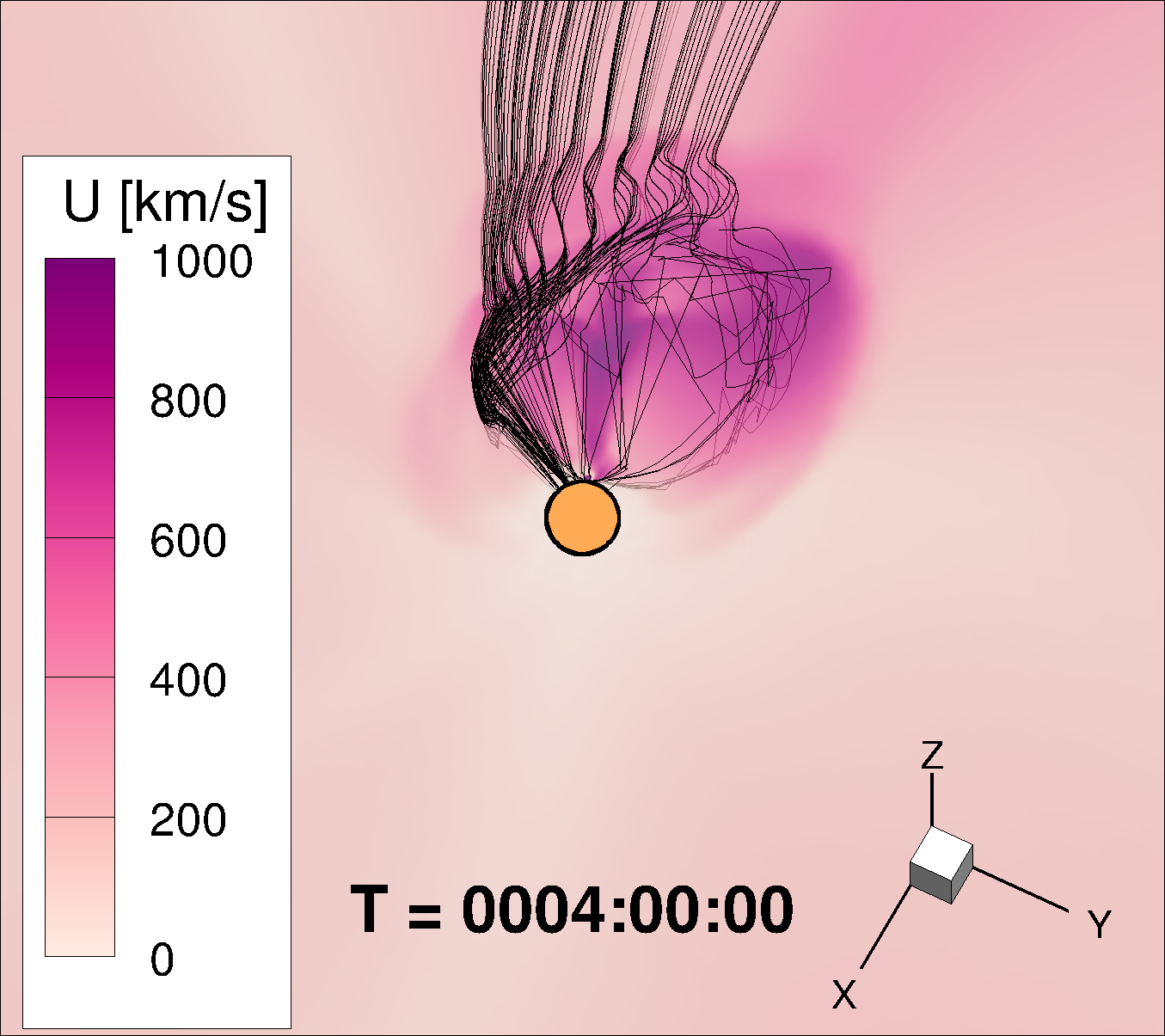}
  \caption{Three snapshots of a CME forming in the corona as seen from
  two angles in HGR: view onto plane of 29.5$^\circ$ latitude ({\bf top}) and plane of 208.5$^\circ$ longitude ({\bf bottom}).
  Color shows the speed of the solar wind.
  Thin black lines are magnetic field lines extracted from the MHD solution and imported to the kinetic code.}
  \label{fig:CME}
\end{figure}

We present some results of the simulation below. 
Figure~\ref{fig:CME} shows three different snapshots 
(2 hours apart) of the CME forming in the solar corona 
together with extracted field lines. 
The CME was initiated
at 29.5$^\circ$ latitude and 208.5$^\circ$ longitude in 
heliographic rotating coordinate system (HGR) and
with anticipated speed $U_{\rm CME}=2000\,{\rm km/s}$.
Figure~\ref{fig:SEPLines} shows SEP flux
for energies exceeding 10~MeV, 
which corresponds to NOAA GOES energy channel~2, along extracted field lines and through 1~AU sphere (interpolated between footprints of filed lines on that sphere).
Finally, Figure~\ref{fig:SEP:GOES} demonstrates a comparison of time evolution of SEP flux with GOES measurements.

\begin{figure}[!t]
  \centering
  \includegraphics[width=0.3\linewidth]{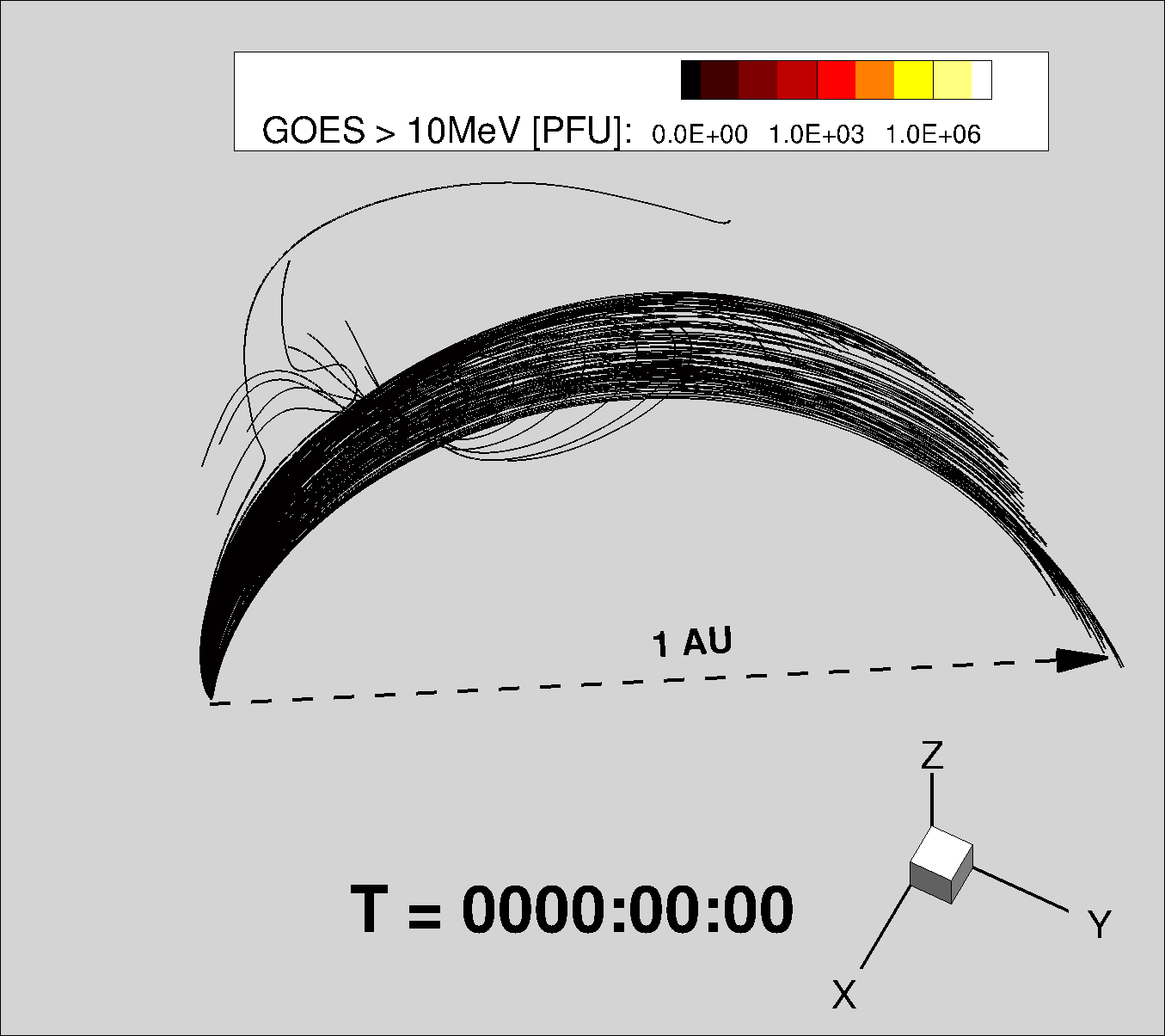}
  \includegraphics[width=0.3\linewidth]{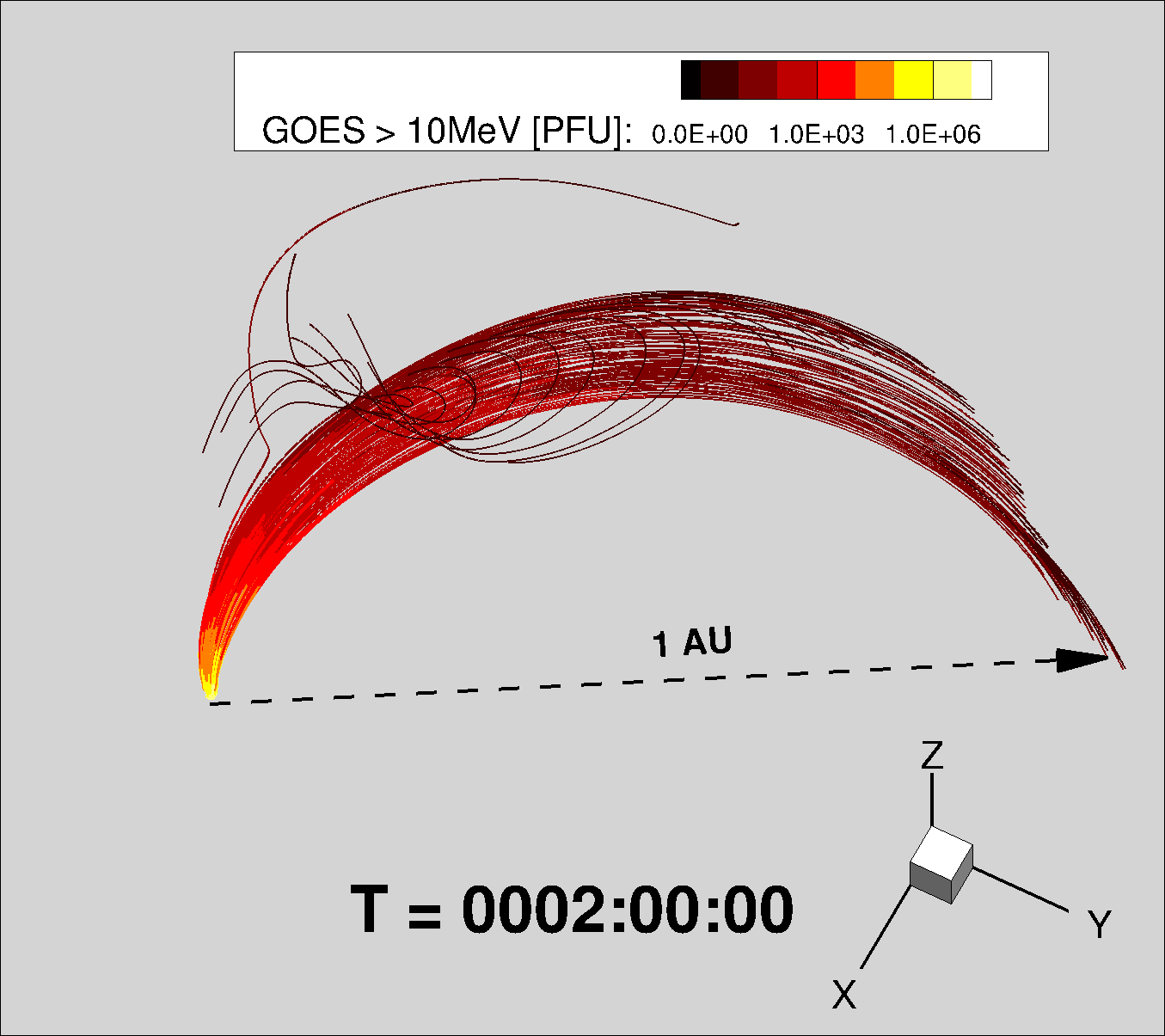}
  \includegraphics[width=0.3\linewidth]{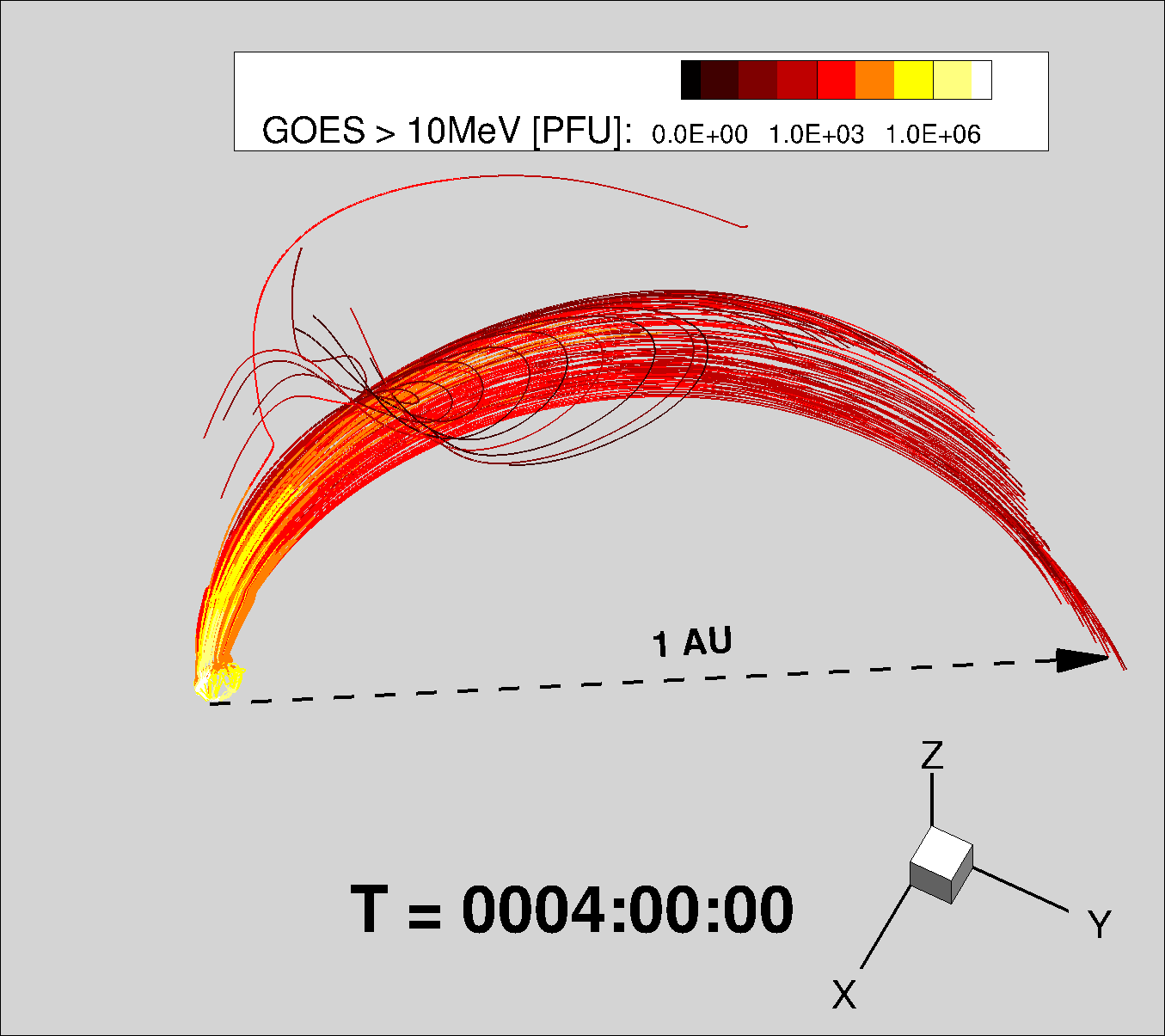}\\
  \includegraphics[width=0.3\linewidth]{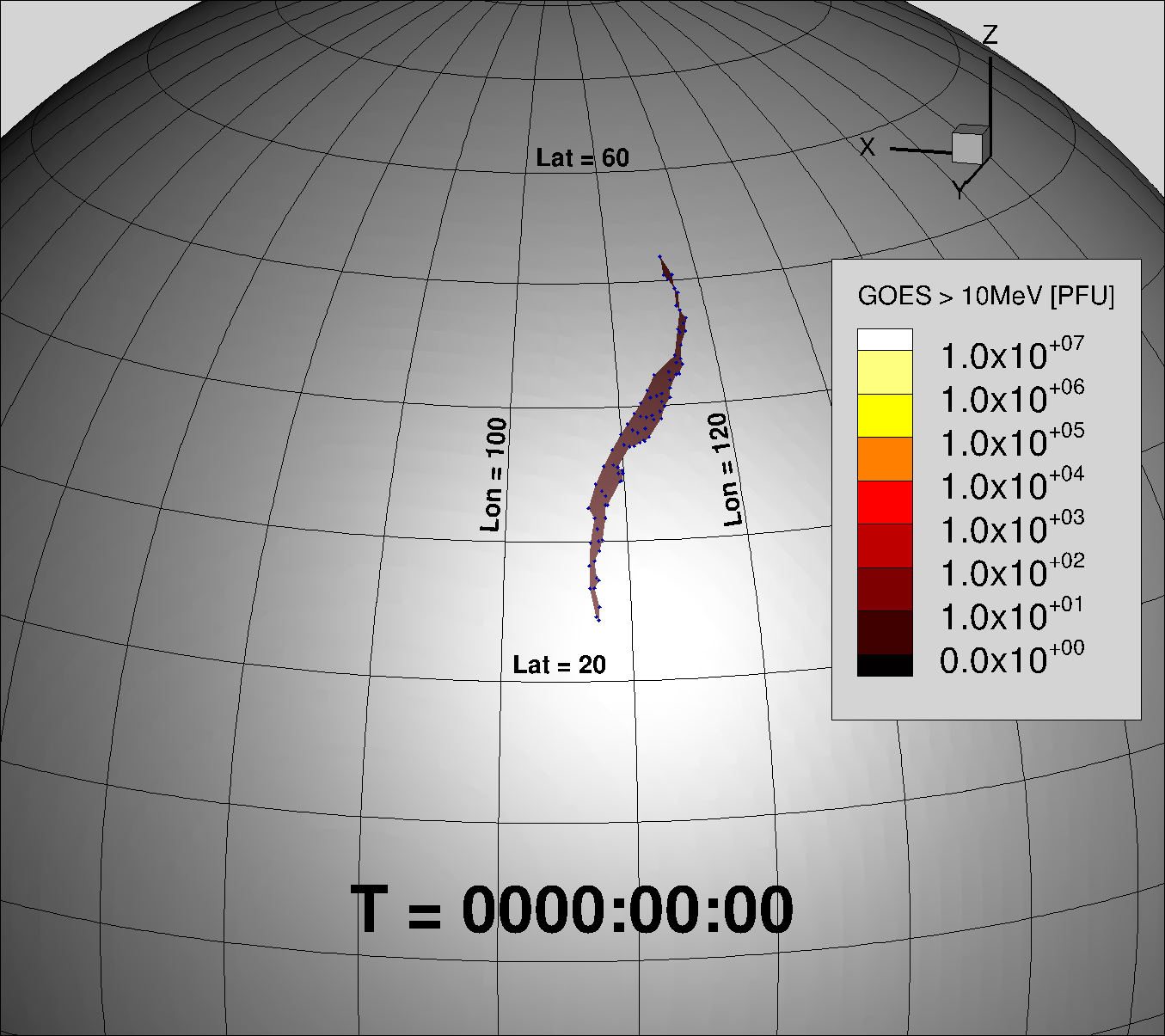}
  \includegraphics[width=0.3\linewidth]{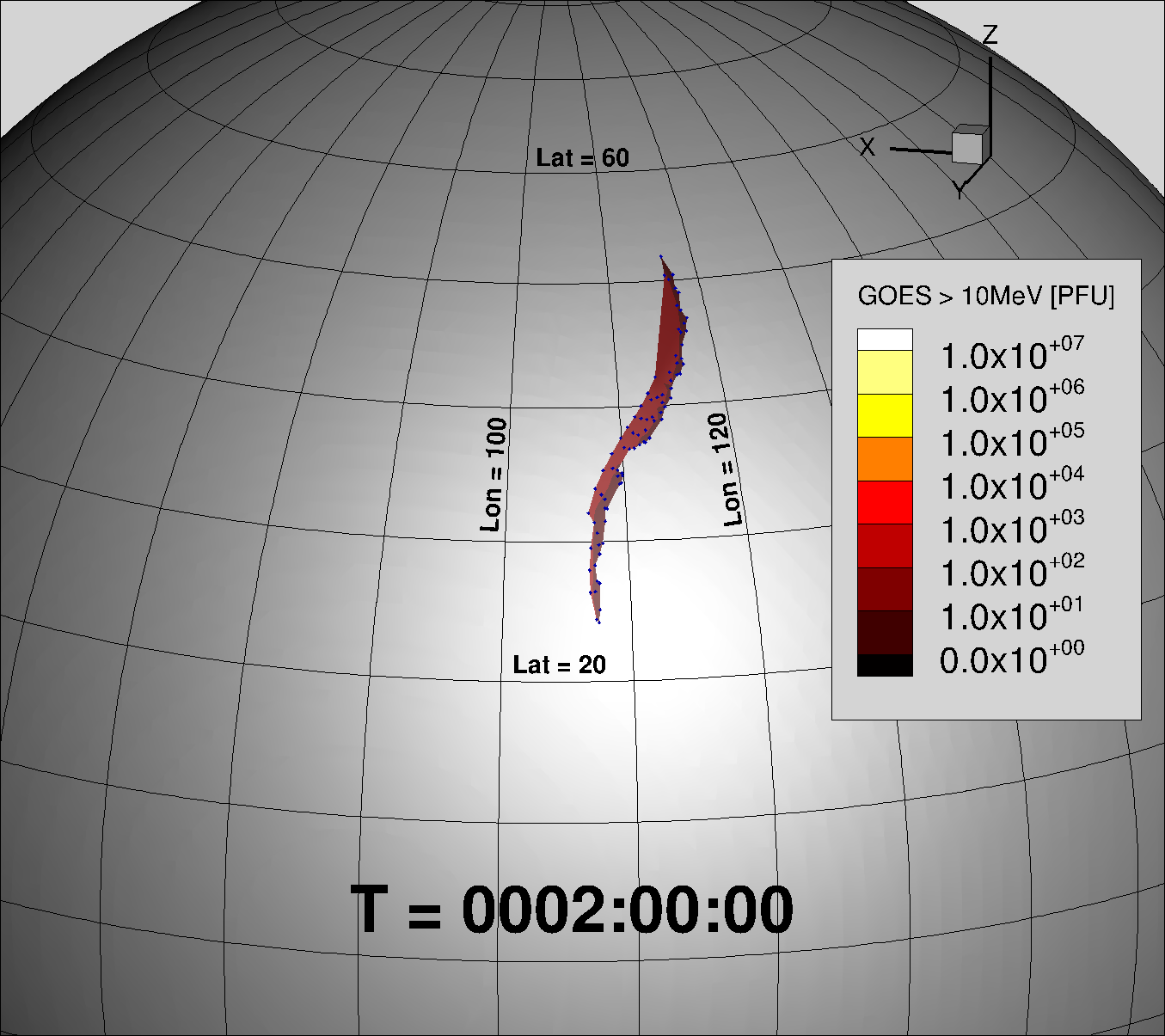}
  \includegraphics[width=0.3\linewidth]{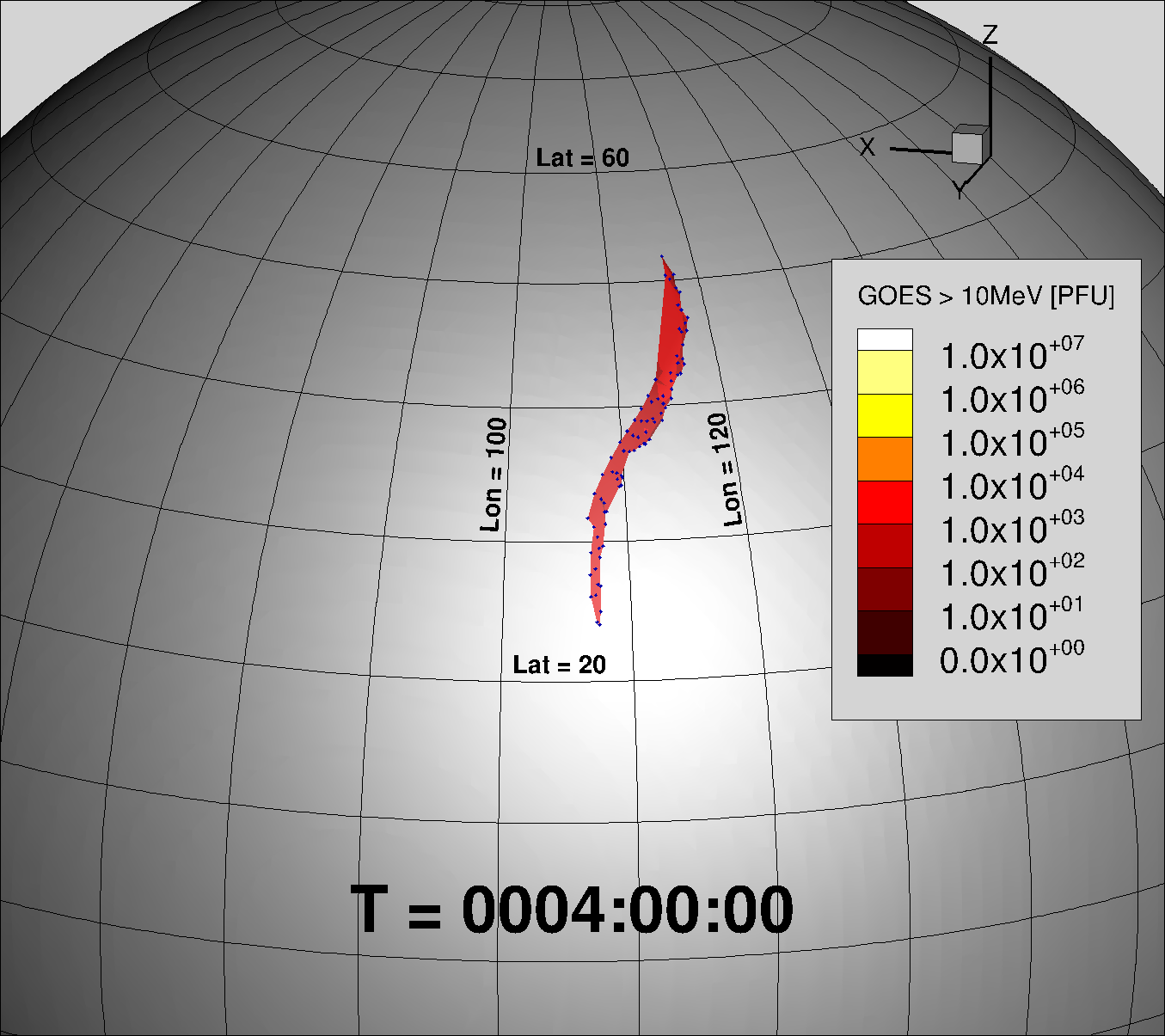}
  \caption{ Simulated flux of SEP exceeding 10~MeV (GOES channel~2): along the extracted lines from the Sun to 1~AU ({\bf top}) and interpolated between footprints (small blue diamonds) of lines on 1~AU sphere ({\bf bottom})}.
  \label{fig:SEPLines}
\end{figure}

\begin{figure}[!t]
  \centering
  \includegraphics[width=0.5\linewidth]{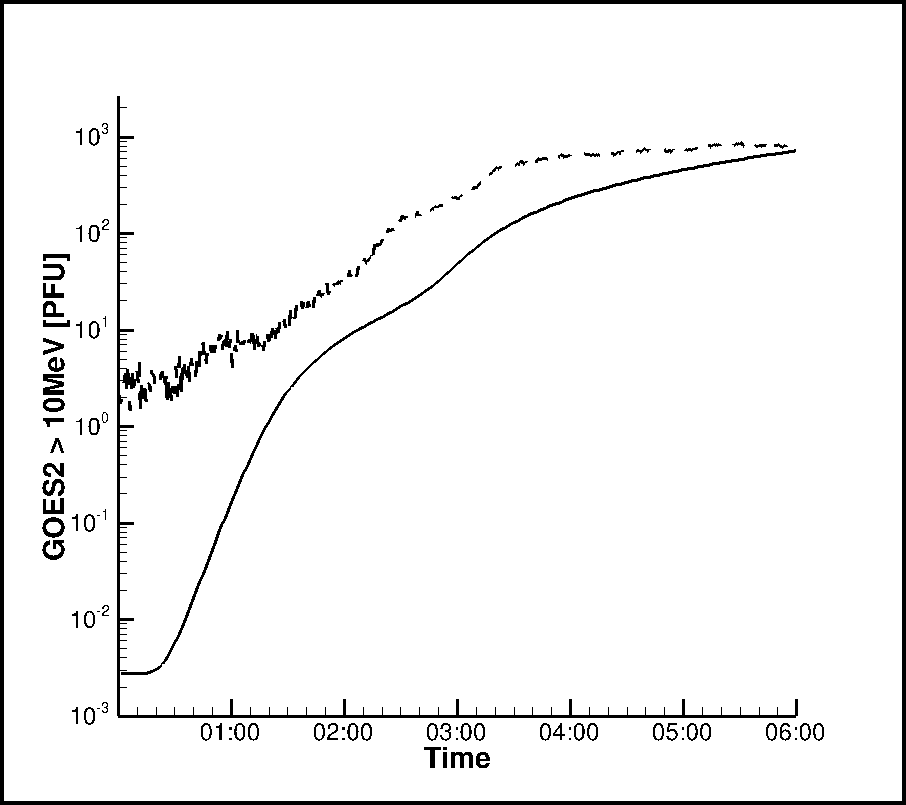}
  \caption{ Time evolution of simulated flux of SEP exceeding 10~MeV (GOES channel~2) at 1~AU on a single line (solid black line) compared to GOES measurements (dashed black line). Time is measured from the CME initiation (4:00 on January 23, 2012).}
  \label{fig:SEP:GOES}
\end{figure}
}

\section{Conclusions}
The present paper serves the purpose of being a guide into and a reference 
for our research effort in the field of SEP forecasting.
In this paper we have reviewed the physical principles that form the basis
of the MHD component of our SEP forecasting framework.
These principles serve as a foundation for a large number of computational models
(developed over the span of several decades)
that allow simulating quiet time SC and IH 
as well as eruptive events in SC and their propagation into IH.
Here, however, we have put more focus on those models and tools
(specifically, AWSoM, AWSoM-R, EEGGL) 
that have been or are being implemented 
in the Space Weather Modeling Framework \citep[SWMF,][]{Toth2012},
which is to host the full SEP forecasting framework.
Additionally, we have suggested the concept of a magnetized cone model,
that could serve as a simple, yet effective eruptive event generator.

The present paper is to be followed by the review of the kinetic component
of our full SEP model and will complete its description.

\section{Acknowledgements}
The collaboration between the CCMC and University of Michigan is supported by the NSF SHINE grant 1257519(PI Aleksandre Taktakishvili). The work performed at the University of Michigan was partially supported by National Science Foundation grants AGS-1322543 and PHY-1513379, NASA grant NNX13AG25G, the European Union’s Horizon 2020 research and innovation program under grant agreement 637302PROGRESS. We would also like to acknowledge high-performance computing support from: (1) Yellowstone(ark:/85065/d7wd3xhc) provided by NCAR’s Computational and Information Systems Laboratory, sponsored by the National Science Foundation,and (2) Pleiades operated by NASA’s Advanced Supercomputing Division.

\appendix

\section{Equilibrium Magnetic Configurations: Spheromak}
\label{sec:Spheromak}
The equations of MHD equilibrium read \citep{landau60}:
\be
\label{eq:GL:start}
{\mathbf{j}}\times{\mathbf{B}} -{\nabla} P = 0,
\ee
which we consider in spherical coordinates 
$\left(r,\theta,\varphi\right)$.
Herewith,
$\mathbf{B}$ is the vector of magnetic field,
$\mathbf{j}$ is the electric current and 
$P$ is the plasma pressure.
It has been demonstrated \citep{grad58,shafranov66}
that an axisymmetric equilibrium MHD configuration 
is governed by  a single scalar equation,
commonly referred to as the Grad-Shafranov equation.
The key concept that allows transforming Eq.~\ref{eq:GL:start}
to this simpler form is that of {\it magnetic surfaces},
which are defined as
surfaces of constant pressure, $P$.

From Eq.~\ref{eq:GL:start}, 
${\mathbf{j}}\cdot{\nabla}P=0$ 
and ${\mathbf{B}}\cdot{\nabla}P=0$,
i.e. a single line of either magnetic field, or electric current
is entirely confined within a single magnetic surface.
Further, magnetic field flux and current  functions defined as
\bea
{\psi}\left({r}_\perp,{z}\right) &=& 
\int_0^{{r}_\perp}{{B}_{z}({r}_\perp^\prime,{z}){r}_\perp^\prime d{r}_\perp^\prime}\nonumber\\
{I}\left({r}_\perp,{z}\right) &=& 
\int_0^{{r}_\perp}{{j}_{z}({r}_\perp^\prime,{z}){r}_\perp^\prime d{r}_\perp^\prime}
\eea
can both be demonstrated to be constant on a given magnetic surface (herewith, ${r}_\perp={r}\sin\theta$ and ${z}={r}\cos\theta$).
Therefore, for an axisymmetric equilibrium configuration, there is a
functional dependence between ${\psi}$, ${I}$ and ${P}$:
${I}{=}{I}({\psi})$, $P{=}P({\psi})$. 
Using Ampere's law, ${\nabla}\times\mathbf{{B}}=\mu_0\mathbf{j}$,
and by introducing the toroidal component of the vector potential, 
$\nabla\times\mathbf{A}=\mathbf{{B}}$, 
one can relate the current and magnetic flux via 
the toroidal components of the field and vector potential: 
$I=\frac{{r}_\perp}{\mu_0}{B}_\varphi$, $\psi={r}_\perp A_\varphi$. 
Thus, the total magnetic field may be expressed as:
\be
\label{eq:GL:B0}
\mathbf{B}=
\nabla\times\left(A_\varphi\mathbf{e}_\varphi\right)+
B_\varphi\mathbf{e}_\varphi=
\frac{1}{{r}_\perp}\left(\nabla\psi\times\mathbf{e}_\varphi+
\mu_0I\mathbf{e}_\varphi
\right)
\ee
Herewith, ${\bf e}_\varphi$ is the unit vector of the azimuthal (toroidal) 
direction.  
Analogously, for the current density vector 
we have: 
\be
\label{eq:GL:j0}
\mu_0\mathbf{j}=
\nabla\times \left[\nabla\times\left(A_\varphi\mathbf{e}_\varphi\right)+{B}_\varphi \mathbf{e}_\varphi\right]=-
\nabla^2\left(A_\varphi\mathbf{e}_\varphi\right)+
\mu_0\frac{dI}{d\psi}
\frac{\nabla\psi\times \mathbf{e}_\varphi}{{r}_\perp}
\ee
Once substitutions Eq.~\ref{eq:GL:B0} and Eq.~\ref{eq:GL:j0} are performed
and a common factor of $\frac{\nabla\psi}{{r}_\perp}$ is omitted,
the condition of equilibrium, Eq.~\ref{eq:GL:start}, reads
\begin{equation}
\label{eq:Shafranov}
\mathbf{e}_\varphi\cdot \nabla^2(A_\varphi{\bf e}_\varphi)=
-\mu_0{r}_\perp \frac{dP}{d\psi}-\mu_0\frac{dI}{d\psi}B_\varphi.
\end{equation}
In the particular case of constant $\frac{dI}{d\psi}$ and $\frac{dP}{d\psi}$,
by expressing the Laplace operator in spherical coordinates Eq.~\ref{eq:Shafranov} reduces to the equation 
describing electro-magnetic waves 
\citep[magnetic dipole and multipole harmonics - see][]{jackson1999}:
\begin{equation}
\label{eq:ShafranovHarm}
\frac1{{r}^2}\frac\partial{\partial{r}}\left({r}^2\frac{\partial A_\varphi}{\partial {r}}\right)
+\frac1{{r}^2\sin\theta}\frac\partial{\partial\theta}\left(\sin\theta\frac{\partial A_\varphi}{\partial\theta}\right)-\frac{A_\varphi}{{r}^2\sin^2\theta}+ \alpha_0^2A_\varphi=-\mu_0 {r}\sin\theta\frac{dP}{d\psi}
\end{equation}
where $\alpha_0=\mu_0 dI/d\psi$.
It may be solved by developing the solution over spherical harmonics: 
$A_\varphi=\sum_{n=1}^\infty{c_nj_n(\alpha_0{r})P^1_n(\cos\theta)} - \mu_0 \alpha_0^{-2}{r}\sin\theta\frac{dP}{d\psi}$,
where $P_n^1$ are associated Legendre polynomials, 
$j_n(x)=\sqrt{\frac\pi{2x}}J_{n+1/2}(x)$,
$J_\nu(x)$ and $j_n(x)$ 
are regular and spherical Bessel functions respectively.
For a dipole harmonic we have: 
\bea
A_\varphi=A_{\varphi0}\left[j_{1}(\alpha_0{r})- \frac{\mu_0{r}}{\alpha_0^{2}A_{\varphi0}}\frac{dP}{d\psi}\right]\sin\theta
\eea 
\begin{figure}
\centering
\includegraphics[width={3in},height={2.5in}]{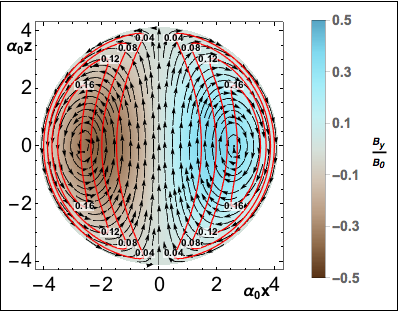}
\includegraphics[width={3in},height={2.5in}]{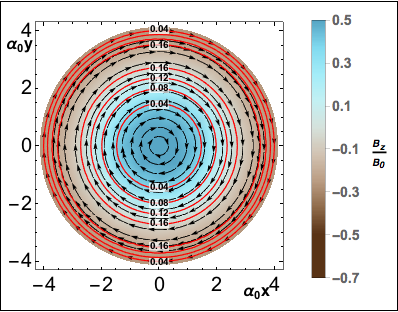}\\
\includegraphics[width={3in}]{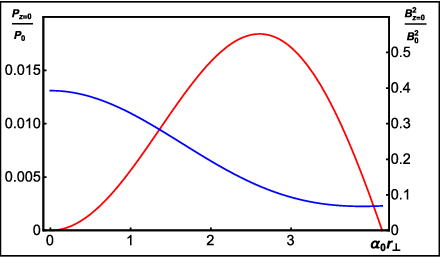}
\includegraphics[width={3in}]{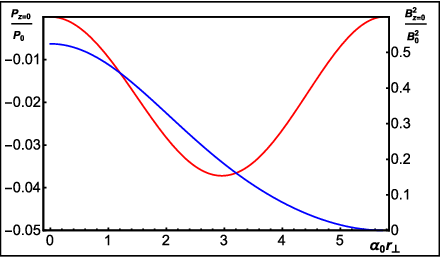}
\caption{\small{
{\bf Top:} spheromak configuration for $\beta_0{=}0.02$:
meridional ({\bf left}) and equatorial ({\bf right}) planes. 
Magnetic field direction is marked with arrows, 
off-plane component of the magnetic field is normalized per $B_0$ 
(see Eq.~\ref{eq:SpheromakBeta}) and shown by color.  
Local values of  plasma parameter 
$\beta({\bf r})=\mu_0P({\bf r})/B^2({\bf r})$ 
are shown with orange curves corresponding to levels 
$\beta=0.04,0.08,0.12,0.16$ as marked explicitly. 
{\bf Bottom}: radial dependence of thermal pressure, $\mu_0P(r)/B^2_0$, (red curve) and 
magnetic pressure, $B^2(r)/B^2_0$, (blue curve) in the equatorial cut ${z}{=}0$: for $\beta_0{=}0.02$
({\bf left panel}) and for 
$\beta_0{=}-2.87{\times}10^{-2}$ ({\bf right panel}) .
}}
\label{Fig:Spheromak}
\end{figure} 

Introducing parameters $B_0=\alpha_0A_{\varphi0}$ and
$\beta_0=\frac{\mu_0}{B_0\alpha_0^2}\frac{dP}{d\psi}$,
we obtain the expression for th spheromak's magnetic field and pressure:
\begin{equation}
\label{eq:app:SpheromakBeta}
{\bf B}_{\rm sk}({\bf r})=\left[\frac{j_1({\alpha_0}{r})}{\alpha_0{r}}-\beta_0\right]\left(2{\bf B}_0+\sigma_h \alpha_0[{\bf B}_0\times{\bf r}]\right)
+j_{2}(\alpha_0{r})\frac{[{\bf r}\times[{\bf r}\times{\bf B}_0]]}{r^2}
\end{equation}
\begin{equation}\label{eq:Pressure}
P_{\rm sk}({\bf r})=\left[\frac{j_1({\alpha_0}{r})}{\alpha_0{r}}-\beta_0\right]\frac{\beta_0 \alpha_0^2[{\bf r}\times{\bf B}_0]^2}{\mu_0}.
\end{equation}
Herewith, the vector ${\bf B}_0$ is introduced with the magnitude equal 
to $B_0$ directed along the polar axis of the spherical coordinate system,
$\sigma_h=\pm1$ is the sign of helicity (we assume $\alpha_0>0$). 
At the center of configuration the magnetic field equals 
${\bf B}_{\rm sk}|_{r=0}=2\left(\frac13-\beta_0\right){\bf B}_0$, 
which for low-beta plasma is only by a numerical factor of ${\approx}0.7$ 
differs from ${\bf B}_0$. 
In Eqs.~\ref{eq:app:SpheromakBeta}-\ref{eq:Pressure}, the coordinate vector, ${\bf r}$, originates at the center of configuration, $\mathbf{R}_{\rm c}$. 
Thus, in the arbitrary coordinate system,
the field and pressure of the configuration equal:  ${\bf B}_{\rm sk}({\bf R}-{\bf R}_{\rm c})$, $p_{\rm sk}({\bf R}-{\bf R}_{\rm c})$, for $\|{\bf R}-{\bf R}_{\rm c}\|\le r_0$. 

We restrict currents to within a spherical magnetic surface 
$\|{\bf R}-{\bf R}_{\rm c}\|=r_0$. 
The radial and toroidal components of the magnetic field turn 
to zero at the surface, thus  $j_1(\alpha_0r_0)=\beta_0\alpha_0r_0$. 
For a given $\beta_0$ this equation relates the configuration size, $r_0$, 
with the extent of magnetic field twisting, $\alpha_0$, needed to close the configuration within this size. 
The plasma pressure, $P$, also turns to zero at the external boundary.

The meridional and equatorial planes  (top) and radial dependence 
of the field and pressure 
for $\beta_0{=}0.02$ (bottom left) are shown 
in Fig.~\ref{Fig:Spheromak}.  
The shown magnetic field lines are also the cross-sections of magnetic surfaces.
\bibliographystyle{authyear}
\bibliography{wholeilr_June8}

\begin{thebibliography}{{\em Hellweg and {Baumstark-Khan}}(2007}

\bibitem[{\em Aguilar et~al.}(2015)]{PRLprotons}
Aguilar, M., D.~Aisa, B.~Alpat, A.~Alvino, G.~Ambrosi, K.~Andeen, L.~Arruda,
  N.~Attig, P.~Azzarello, A.~Bachlechner, F.~Barao, A.~Barrau, L.~Barrin,
  A.~Bartoloni, L.~Basara, M.~Battarbee, R.~Battiston, J.~Bazo, U.~Becker,
  M.~Behlmann, B.~Beischer, J.~Berdugo, B.~Bertucci, G.~Bigongiari, V.~Bindi,
  S.~Bizzaglia, M.~Bizzarri, G.~Boella, W.~de~Boer, K.~Bollweg, V.~Bonnivard,
  B.~Borgia, S.~Borsini, M.~J. Boschini, M.~Bourquin, J.~Burger, F.~Cadoux,
  X.~D. Cai, M.~Capell, S.~Caroff, J.~Casaus, V.~Cascioli, G.~Castellini,
  I.~Cernuda, D.~Cerreta, F.~Cervelli, M.~J. Chae, Y.~H. Chang, A.~I. Chen,
  H.~Chen, G.~M. Cheng, H.~S. Chen, L.~Cheng, H.~Y. Chou, E.~Choumilov,
  V.~Choutko, C.~H. Chung, C.~Clark, R.~Clavero, G.~Coignet, C.~Consolandi,
  A.~Contin, C.~Corti, E.~Cortina Gil, B.~Coste, W.~Creus, M.~Crispoltoni,
  Z.~Cui, Y.~M. Dai, C.~Delgado, S.~Della~Torre, M.~B. Demirk\"oz, L.~Derome,
  S.~Di~Falco, L.~Di~Masso, F.~Dimiccoli, C.~D\'{\i}az, P.~von Doetinchem,
  F.~Donnini, W.~J. Du, M.~Duranti, D.~D'Urso, A.~Eline, F.~J. Eppling,
  T.~Eronen, Y.~Y. Fan, L.~Farnesini, J.~Feng, E.~Fiandrini, A.~Fiasson,
  E.~Finch, P.~Fisher, Y.~Galaktionov, G.~Gallucci, B.~Garc\'{\i}a,
  R.~Garc\'{\i}a-L\'opez, C.~Gargiulo, H.~Gast, I.~Gebauer, M.~Gervasi,
  A.~Ghelfi, W.~Gillard, F.~Giovacchini, P.~Goglov, J.~Gong, C.~Goy,
  V.~Grabski, D.~Grandi, M.~Graziani, C.~Guandalini, I.~Guerri, K.~H. Guo,
  D.~Haas, M.~Habiby, S.~Haino, K.~C. Han, Z.~H. He, M.~Heil, J.~Hoffman, T.~H.
  Hsieh, Z.~C. Huang, C.~Huh, M.~Incagli, M.~Ionica, W.~Y. Jang, H.~Jinchi,
  K.~Kanishev, G.~N. Kim, K.~S. Kim, Th. Kirn, R.~Kossakowski, O.~Kounina,
  A.~Kounine, V.~Koutsenko, M.~S. Krafczyk, G.~La~Vacca, E.~Laudi, G.~Laurenti,
  I.~Lazzizzera, A.~Lebedev, H.~T. Lee, S.~C. Lee, C.~Leluc, G.~Levi, H.~L. Li,
  J.~Q. Li, Q.~Li, Q.~Li, T.~X. Li, W.~Li, Y.~Li, Z.~H. Li, Z.~Y. Li, S.~Lim,
  C.~H. Lin, P.~Lipari, T.~Lippert, D.~Liu, H.~Liu, M.~Lolli, T.~Lomtadze,
  M.~J. Lu, S.~Q. Lu, Y.~S. Lu, K.~Luebelsmeyer, J.~Z. Luo, S.~S. Lv, R.~Majka,
  C.~Ma\~n\'a, J.~Mar\'{\i}n, T.~Martin, G.~Mart\'{\i}nez, N.~Masi, D.~Maurin,
  A.~Menchaca-Rocha, Q.~Meng, D.~C. Mo, L.~Morescalchi, P.~Mott, M.~M\"uller,
  J.~Q. Ni, N.~Nikonov, F.~Nozzoli, P.~Nunes, A.~Obermeier, A.~Oliva,
  M.~Orcinha, F.~Palmonari, C.~Palomares, M.~Paniccia, A.~Papi, M.~Pauluzzi,
  E.~Pedreschi, S.~Pensotti, R.~Pereira, N.~Picot-Clemente, F.~Pilo, A.~Piluso,
  C.~Pizzolotto, V.~Plyaskin, M.~Pohl, V.~Poireau, E.~Postaci, A.~Putze,
  L.~Quadrani, X.~M. Qi, X.~Qin, Z.~Y. Qu, T.~R\"aih\"a, P.~G. Rancoita,
  D.~Rapin, J.~S. Ricol, I.~Rodr\'{\i}guez, S.~Rosier-Lees, A.~Rozhkov,
  D.~Rozza, R.~Sagdeev, J.~Sandweiss, P.~Saouter, C.~Sbarra, S.~Schael, S.~M.
  Schmidt, A.~Schulz von Dratzig, G.~Schwering, G.~Scolieri, E.~S. Seo, B.~S.
  Shan, Y.~H. Shan, J.~Y. Shi, X.~Y. Shi, Y.~M. Shi, T.~Siedenburg, D.~Son,
  F.~Spada, F.~Spinella, W.~Sun, W.~H. Sun, M.~Tacconi, C.~P. Tang, X.~W. Tang,
  Z.~C. Tang, L.~Tao, D.~Tescaro, Samuel C.~C. Ting, S.~M. Ting, N.~Tomassetti,
  J.~Torsti, C.~T\"urko\ifmmode~\breve{g}\else \u{g}\fi{}lu, T.~Urban,
  V.~Vagelli, E.~Valente, C.~Vannini, E.~Valtonen, S.~Vaurynovich, M.~Vecchi,
  M.~Velasco, J.~P. Vialle, V.~Vitale, S.~Vitillo, L.~Q. Wang, N.~H. Wang,
  Q.~L. Wang, R.~S. Wang, X.~Wang, Z.~X. Wang, Z.~L. Weng, K.~Whitman,
  J.~Wienkenh\"over, H.~Wu, X.~Wu, X.~Xia, M.~Xie, S.~Xie, R.~Q. Xiong, G.~M.
  Xin, N.~S. Xu, W.~Xu, Q.~Yan, J.~Yang, M.~Yang, Q.~H. Ye, H.~Yi, Y.~J. Yu,
  Z.~Q. Yu, S.~Zeissler, J.~H. Zhang, M.~T. Zhang, X.~B. Zhang, Z.~Zhang, Z.~M.
  Zheng, H.~L. Zhuang, V.~Zhukov, A.~Zichichi, N.~Zimmermann, P.~Zuccon, and
  C.~Zurbach,
\newblock Precision measurement of the proton flux in primary cosmic rays from
  rigidity 1 gv to 1.8 tv with the alpha magnetic spectrometer on the
  international space station,
\newblock {\em Phys. Rev. Lett.}, {\em 114}, 171103, Apr 2015.

\bibitem[{\em {Alazraki} and {Couturier}}(1971)]{alazraki71}
{Alazraki}, G., and P.~{Couturier},
\newblock {Solar Wind Acceleration Caused by the Gradient of Alfv\'en Wave
  Pressure},
\newblock {\em Astron. \& Astrophys.}, {\em 13}, 380, August 1971.

\bibitem[{\em {Alfv{\'e}n}}(1942)]{Alfven1942}
{Alfv{\'e}n}, H.,
\newblock {Existence of Electromagnetic-Hydrodynamic Waves},
\newblock {\em \nat}, {\em 150}, 405--406, October 1942.

\bibitem[{\em {Antiochos} et~al.}(1999)]{Antiochos1999}
{Antiochos}, S.~K., C.~R. {DeVore}, and J.~A. {Klimchuk},
\newblock {A Model for Solar Coronal Mass Ejections},
\newblock {\em \apj}, {\em 510}, 485--493, January 1999.

\bibitem[{\em {Arge} and {Pizzo}}(2000)]{Arge2000}
{Arge}, C.~N., and V.~J. {Pizzo},
\newblock {Improvement in the prediction of solar wind conditions using
  near-real time solar magnetic field updates},
\newblock {\em \jgr}, {\em 105}, 10465--10480, May 2000.

\bibitem[{\em {Arge} et~al.}(2003)]{Arge2003}
{Arge}, C.~N., D.~{Odstrcil}, V.~J. {Pizzo}, and L.~R. {Mayer},
\newblock {Improved Method for Specifying Solar Wind Speed Near the Sun},
\newblock In {Velli}, M., R.~{Bruno}, F.~{Malara}, and B.~{Bucci}, editors,
  {\em Solar Wind Ten}, volume 679 of {\em American Institute of Physics
  Conference Series}, pages 190--193, September 2003.

\bibitem[{\em {Axford} et~al.}(1977)]{Axford1977}
{Axford}, W.~I., E.~{Leer}, and G.~{Skadron},
\newblock {The acceleration of cosmic rays by shock waves},
\newblock {\em International Cosmic Ray Conference}, {\em 11}, 132--137, 1977.

\bibitem[{\em {Barnes}}(1966)]{Barnes1966}
{Barnes}, A.,
\newblock {Collisionless Damping of Hydromagnetic Waves},
\newblock {\em Physics of Fluids}, {\em 9}, 1483--1495, August 1966.

\bibitem[{\em {Barnes}}(1968)]{Barnes1968}
{Barnes}, A.,
\newblock {Collisionless Heating of the Solar-Wind Plasma. I. Theory of the
  Heating of Collisionless Plasma by Hydromagnetic Waves},
\newblock {\em \apj}, {\em 154}, 751, November 1968.

\bibitem[{\em {Belcher} and {Davis}}(1971)]{Belcher1971}
{Belcher}, J.~W., and L.~{Davis}, Jr.,
\newblock {Large-amplitude Alfv{\'e}n waves in the interplanetary medium, 2},
\newblock {\em \jgr}, {\em 76}, 3534, 1971.

\bibitem[{\em {Belcher} et~al.}(1969)]{Belcher1969}
{Belcher}, J.~W., L.~{Davis}, Jr., and E.~J. {Smith},
\newblock {Large-amplitude Alfv{\'e}n waves in the interplanetary medium:
  Mariner 5},
\newblock {\em \jgr}, {\em 74}, 2302, 1969.

\bibitem[{\em {Bell}}(1978a)]{Bell1978a}
{Bell}, A.~R.,
\newblock {The acceleration of cosmic rays in shock fronts. I},
\newblock {\em \mnras}, {\em 182}, 147--156, January 1978a.

\bibitem[{\em {Bell}}(1978b)]{Bell1978b}
{Bell}, A.~R.,
\newblock {The acceleration of cosmic rays in shock fronts. II},
\newblock {\em \mnras}, {\em 182}, 443--455, February 1978b.

\bibitem[{\em {Blandford} and {Ostriker}}(1978)]{Blandford1978}
{Blandford}, R.~D., and J.~P. {Ostriker},
\newblock {Particle acceleration by astrophysical shocks},
\newblock {\em \apjl}, {\em 221}, L29--L32, April 1978.

\bibitem[{\em Borovikov et~al.}(2015)]{Borovikov2015}
Borovikov, Dmitry, Igor~V. Sokolov, and G{\'{a}}bor T{\'{o}}th,
\newblock An efficient second-order accurate and continuous interpolation for
  block-adaptive grids,
\newblock {\em J. Comput. Physics}, {\em 297}, 599--610, 2015.

\bibitem[{\em {Borovikov} et~al.}(2017)]{borovikov17}
{Borovikov}, D., I.~V. {Sokolov}, W.~B. {Manchester}, M.~{Jin}, and T.~I.
  {Gombosi},
\newblock {Eruptive event generator based on the Gibson-Low magnetic
  configuration},
\newblock {\em Journal of Geophysical Research (Space Physics)}, {\em 122},
  7979--7984, August 2017.

\bibitem[{\em {Brueckner} et~al.}(1995)]{brueckner95}
{Brueckner}, G.~E., R.~A. {Howard}, M.~J. {Koomen}, C.~M. {Korendyke}, D.~J.
  {Michels}, J.~D. {Moses}, D.~G. {Socker}, K.~P. {Dere}, P.~L. {Lamy},
  A.~{Llebaria}, M.~V. {Bout}, R.~{Schwenn}, G.~M. {Simnett}, D.~K. {Bedford},
  and C.~J. {Eyles},
\newblock {The Large Angle Spectroscopic Coronagraph (LASCO)},
\newblock {\em \solphys}, {\em 162}, 357--402, December 1995.

\bibitem[{\em {Cane} et~al.}(2006)]{cane06}
{Cane}, H.~V., R.~A. {Mewaldt}, C.~M.~S. {Cohen}, and T.~T. {von Rosenvinge},
\newblock {Role of flares and shocks in determining solar energetic particle
  abundances},
\newblock {\em Journal of Geophysical Research (Space Physics)}, {\em 111},
  A06S90, June 2006.

\bibitem[{\em {Cliver} et~al.}(2004)]{Cliver2004}
{Cliver}, E.~W., S.~W. {Kahler}, and D.~V. {Reames},
\newblock {Coronal Shocks and Solar Energetic Proton Events},
\newblock {\em \apj}, {\em 605}, 902--910, April 2004.

\bibitem[{\em Cliver}(2006)]{Cliver06}
Cliver, E.~W.,
\newblock The 1859 space weather event: Then and now,
\newblock {\em Adv. Space Res.}, {\em 38}( 2 ), 119--129, 2006.

\bibitem[{\em {Cliver}}(2009)]{cliver09}
{Cliver}, E.~W.,
\newblock {History of research on solar energetic particle (SEP) events: the
  evolving paradigm},
\newblock In {Gopalswamy}, N., and D.~F. {Webb}, editors, {\em Universal
  Heliophysical Processes}, volume 257 of {\em IAU Symposium}, pages 401--412,
  March 2009.

\bibitem[{\em {Cohen} et~al.}(1999)]{cohen99}
{Cohen}, C.~M.~S., R.~A. {Mewaldt}, R.~A. {Leske}, A.~C. {Cummings}, E.~C.
  {Stone}, M.~E. {Wiedenbeck}, E.~R. {Christian}, and T.~T. {von Rosenvinge},
\newblock {New observations of heavy-ion-rich solar particle events from ACE},
\newblock {\em \grl}, {\em 26}, 2697--2700, 1999.

\bibitem[{\em {Cohen} et~al.}(2007)]{cohen07}
{Cohen}, O., I.~V. {Sokolov}, I.~I. {Roussev}, C.~N. {Arge}, W.~B.
  {Manchester}, T.~I. {Gombosi}, R.~A. {Frazin}, H.~{Park}, M.~D. {Butala},
  F.~{Kamalabadi}, and M.~{Velli},
\newblock A semiempirical magnetohydrodynamical model of the solar wind,
\newblock {\em Astrophys. J. Lett.}, {\em 654}, L163--L166, January 2007.

\bibitem[{\em {Cohen} et~al.}(2008)]{cohen08}
{Cohen}, O., I.~V. {Sokolov}, I.~I. {Roussev}, and T.~I. {Gombosi},
\newblock Validation of a synoptic solar wind model,
\newblock {\em J. Geophys. Res.}, {\em 113}( A12 ), A03104, March 2008.

\bibitem[{\em {Coleman}}(1966)]{Coleman1966}
{Coleman}, P.~J., Jr.,
\newblock {Variations in the interplanetary magnetic field: Mariner 2: 1.
  Observed properties},
\newblock {\em \jgr}, {\em 71}, 5509--5531, December 1966.

\bibitem[{\em {Coleman}}(1967)]{Coleman1967}
{Coleman}, P.~J., Jr.,
\newblock {Wave-like phenomena in the interplanetary plasma: Mariner 2},
\newblock {\em \planss}, {\em 15}, 953--973, June 1967.

\bibitem[{\em {Coleman}}(1968)]{Coleman1968}
{Coleman}, P.~J., Jr.,
\newblock {Turbulence, Viscosity, and Dissipation in the Solar-Wind Plasma},
\newblock {\em \apj}, {\em 153}, 371, August 1968.

\bibitem[{\em {Cranmer}}(2010)]{Cranmer2010}
{Cranmer}, S.~R.,
\newblock {An Efficient Approximation of the Coronal Heating Rate for use in
  Global Sun-Heliosphere Simulations},
\newblock {\em \apj}, {\em 710}, 676--688, February 2010.

\bibitem[{\em {Dewar}}(1970)]{Dewar1970}
{Dewar}, R.~L.,
\newblock {Interaction between Hydromagnetic Waves and a Time-Dependent,
  Inhomogeneous Medium},
\newblock {\em Physics of Fluids}, {\em 13}, 2710--2720, November 1970.

\bibitem[{\em {Dmitruk} et~al.}(2002)]{Dmitruk2002}
{Dmitruk}, P., W.~H. {Matthaeus}, L.~J. {Milano}, S.~{Oughton}, G.~P. {Zank},
  and D.~J. {Mullan},
\newblock {Coronal Heating Distribution Due to Low-Frequency, Wave-driven
  Turbulence},
\newblock {\em \apj}, {\em 575}, 571--577, August 2002.

\bibitem[{\em {Downs} et~al.}(2010)]{Downs2010}
{Downs}, C., I.~I. {Roussev}, B.~{van der Holst}, N.~{Lugaz}, I.~V. {Sokolov},
  and T.~I. {Gombosi},
\newblock {Toward a Realistic Thermodynamic Magnetohydrodynamic Model of the
  Global Solar Corona},
\newblock {\em \apj}, {\em 712}, 1219--1231, April 2010.

\bibitem[{\em {Els{\"a}sser}}(1950)]{Elsasser1950}
{Els{\"a}sser}, W.~M.,
\newblock {The Hydromagnetic Equations},
\newblock {\em Physical Review}, {\em 79}, 183--183, July 1950.

\bibitem[{\em Fan and Gibson}(2004)]{Fan2004}
Fan, Y., and S.~E. Gibson,
\newblock Numerical simulations of three-dimensional coronal magnetic fields
  resulting from the emergence of twisted magnetic flux tubes,
\newblock {\em The Astrophysical Journal}, {\em 609}( 2 ), 1123, 2004.

\bibitem[{\em Fermi}(1949)]{fermi49}
Fermi, E.,
\newblock On the origin of the cosmic radiation,
\newblock {\em Phys. Rev.}, {\em 75}, 1169--1174, Apr 1949.

\bibitem[{\em {Forbush}}(1946)]{forbush46}
{Forbush}, S.~E.,
\newblock {Three Unusual Cosmic-Ray Increases Possibly Due to Charged Particles
  from the Sun},
\newblock {\em Physical Review}, {\em 70}, 771--772, November 1946.

\bibitem[{\em {Gibson} and {Low}}(1998)]{gibson98}
{Gibson}, S.~E., and B.~C. {Low},
\newblock {A Time-Dependent Three-Dimensional Magnetohydrodynamic Model of the
  Coronal Mass Ejection},
\newblock {\em Astrophys. J.}, {\em 493}, 460--473, January 1998.

\bibitem[{\em {Gopalswamy} et~al.}(2001)]{Gopalswamy2001}
{Gopalswamy}, N., A.~{Lara}, M.~L. {Kaiser}, and J.-L. {Bougeret},
\newblock {Near-Sun and near-Earth manifestations of solar eruptions},
\newblock {\em \jgr}, {\em 106}, 25261--25278, November 2001.

\bibitem[{\em {Gopalswamy} et~al.}(2012)]{gopalswamy12}
{Gopalswamy}, N., H.~{Xie}, S.~{Yashiro}, S.~{Akiyama}, P.~{M{\"a}kel{\"a}},
  and I.~G. {Usoskin},
\newblock {Properties of Ground Level Enhancement Events and the Associated
  Solar Eruptions During Solar Cycle 23},
\newblock {\em \ssr}, {\em 171}, 23--60, October 2012.

\bibitem[{\em {Gopalswamy} et~al.}(2014)]{gopalswamy14}
{Gopalswamy}, N., H.~{Xie}, S.~{Akiyama}, P.~A. {M{\"a}kel{\"a}}, and
  S.~{Yashiro},
\newblock {Major solar eruptions and high-energy particle events during solar
  cycle 24},
\newblock {\em Earth, Planets, and Space}, {\em 66}, 104, December 2014.

\bibitem[{\em {Gosling}}(1993)]{Gosling1993}
{Gosling}, J.~T.,
\newblock {The solar flare myth},
\newblock {\em \jgr}, {\em 98}, 18937--18950, November 1993.

\bibitem[{\em {Grad} and {Rubin}}(1958)]{grad58}
{Grad}, H., and H~{Rubin},
\newblock {Hydromagnetic Equilibria and Force-Free Fields},
\newblock In {\em Proceedings of the 2nd UN Conference on the Peaceful Uses of
  Atomic Energy}, volume~31, pages 190--197, 1958.

\bibitem[{\em {Grechnev} et~al.}(2008)]{grechnev08}
{Grechnev}, V.~V., V.~G. {Kurt}, I.~M. {Chertok}, A.~M. {Uralov},
  H.~{Nakajima}, A.~T. {Altyntsev}, A.~V. {Belov}, B.~Y. {Yushkov}, S.~N.
  {Kuznetsov}, L.~K. {Kashapova}, N.~S. {Meshalkina}, and N.~P. {Prestage},
\newblock {An Extreme Solar Event of 20 January 2005: Properties of the Flare
  and the Origin of Energetic Particles},
\newblock {\em \solphys}, {\em 252}, 149--177, October 2008.

\bibitem[{\em {Green} and {Boardsen}}(2006)]{Green06a}
{Green}, J.~L., and S.~{Boardsen},
\newblock Duration and extent of the great auroral storm of 1859,
\newblock {\em Adv. Space Res.}, {\em 38}( 2 ), 130--135, 2006.

\bibitem[{\em {Green} et~al.}(2006)]{Green06}
{Green}, James~L., Scott {Boardsen}, Sten {Odenwald}, John {Humble}, and
  Katherine~A. {Pazamickas},
\newblock Eyewitness reports of the great auroral storm of 1859,
\newblock {\em Adv. Space Res.}, {\em 38}( 2 ), 145--154, 2006.

\bibitem[{\em {Gressl} et~al.}(2014)]{Gressl2014}
{Gressl}, C., A.~M. {Veronig}, M.~{Temmer}, D.~{Odstr{\v c}il}, J.~A. {Linker},
  Z.~{Miki{\'c}}, and P.~{Riley},
\newblock {Comparative Study of MHD Modeling of the Background Solar Wind},
\newblock {\em \solphys}, {\em 289}, 1783--1801, May 2014.

\bibitem[{\em {Heath} et~al.}(1977)]{Heath77}
{Heath}, D.~F., A.~J. {Krueger}, and P.~J. {Crutzen},
\newblock Solar proton event: influence on stratospheric ozone,
\newblock {\em Science}, {\em 197}( 4306 ), 886--9, 1977.

\bibitem[{\em {Heinemann} and {Olbert}}(1980)]{Heinemann1980}
{Heinemann}, M., and S.~{Olbert},
\newblock {Non-WKB Alfven waves in the solar wind},
\newblock {\em \jgr}, {\em 85}, 1311--1327, March 1980.

\bibitem[{\em Hellweg and {Baumstark-Khan}}(2007)]{Hellweg:2007}
Hellweg, C.~E., and C.~{Baumstark-Khan},
\newblock Getting ready for the manned mission to {Mars}: The astronauts' risk
  from space radiation,
\newblock {\em Naturwissenschaften}, {\em 94}( 7 ), 517--526, jan 2007.

\bibitem[{\em {Hollweg}}(1986)]{Hollweg1986}
{Hollweg}, J.~V.,
\newblock {Transition region, corona, and solar wind in coronal holes},
\newblock {\em \jgr}, {\em 91}, 4111--4125, April 1986.

\bibitem[{\em {Hu} et~al.}(2003)]{Hu2003}
{Hu}, Y.~Q., S.~R. {Habbal}, Y.~{Chen}, and X.~{Li},
\newblock {Are coronal holes the only source of fast solar wind at solar
  minimum?},
\newblock {\em Journal of Geophysical Research (Space Physics)}, {\em 108},
  1377, October 2003.

\bibitem[{\em Jackson}(1999)]{jackson1999}
Jackson, John~David,
\newblock {\em Classical electrodynamics},
\newblock Wiley, New York, {NY}, 3rd ed. edition, 1999.

\bibitem[{\em {Jacques}}(1977)]{Jacques1977}
{Jacques}, S.~A.,
\newblock {Momentum and energy transport by waves in the solar atmosphere and
  solar wind},
\newblock {\em \apj}, {\em 215}, 942--951, August 1977.

\bibitem[{\em {Jacques}}(1978)]{Jacques1978}
{Jacques}, S.~A.,
\newblock {Solar wind models with Alfven waves},
\newblock {\em \apj}, {\em 226}, 632--649, December 1978.

\bibitem[{\em J{\"a}kel}(2004)]{Jakel:2004}
J{\"a}kel, O.,
\newblock Radiation hazard during a manned mission to {Mars},
\newblock {\em Zeitschrift fur medizinische Physik}, {\em 14}( 4 ), 267--272,
  2004.

\bibitem[{\em {Jian} et~al.}(2015)]{Jian2015}
{Jian}, L.~K., P.~J. {MacNeice}, A.~{Taktakishvili}, D.~{Odstrcil},
  B.~{Jackson}, H.-S. {Yu}, P.~{Riley}, I.~V. {Sokolov}, and R.~M. {Evans},
\newblock {Validation for solar wind prediction at Earth: Comparison of coronal
  and heliospheric models installed at the CCMC},
\newblock {\em Space Weather}, {\em 13}, 316--338, May 2015.

\bibitem[{\em {Jin} et~al.}(2013)]{Jin2013}
{Jin}, M., W.~B. {Manchester}, B.~{van der Holst}, R.~{Oran}, I.~{Sokolov},
  G.~{Toth}, Y.~{Liu}, X.~D. {Sun}, and T.~I. {Gombosi},
\newblock {Numerical Simulations of Coronal Mass Ejection on 2011 March 7:
  One-temperature and Two-temperature Model Comparison},
\newblock {\em \apj}, {\em 773}, 50, August 2013.

\bibitem[{\em Jin et~al.}(2017a)]{Jin:2017a}
Jin, M., W.~B. Manchester, B.~van~der Holst, I.~Sokolov, G.~T{\'o}th, R.~E.
  Mullinix, A.~Taktakishvili, A.~Chulaki, and T.~I. Gombosi,
\newblock Data-constrained coronal mass ejections in a global
  magnetohydrodynamics model,
\newblock {\em The Astrophysical Journal}, {\em 834}( 2 ), 173, 2017a.

\bibitem[{\em Jin et~al.}(2017b)]{Jin:2017b}
Jin, M., W.~B. Manchester, B.~van~der Holst, I.~Sokolov, G.~T{\'o}th,
  A.~Vourlidas, C.~A. de~Koning, and T.~I. Gombosi,
\newblock Chromosphere to 1 au simulation of the 2011 march 7th event: A
  comprehensive study of coronal mass ejection propagation,
\newblock {\em The Astrophysical Journal}, {\em 834}( 2 ), 172, 2017b.

\bibitem[{\em {Joshi} et~al.}(2013)]{joshi13}
{Joshi}, N.~C., W.~{Uddin}, A.~K. {Srivastava}, R.~{Chandra}, N.~{Gopalswamy},
  P.~K. {Manoharan}, M.~J. {Aschwanden}, D.~P. {Choudhary}, R.~{Jain}, N.~V.
  {Nitta}, H.~{Xie}, S.~{Yashiro}, S.~{Akiyama}, P.~{M{\"a}kel{\"a}},
  P.~{Kayshap}, A.~K. {Awasthi}, V.~C. {Dwivedi}, and K.~{Mahalakshmi},
\newblock {A multiwavelength study of eruptive events on January 23, 2012
  associated with a major solar energetic particle event},
\newblock {\em Advances in Space Research}, {\em 52}, 1--14, July 2013.

\bibitem[{\em {Joyce} et~al.}(2015)]{joyce15}
{Joyce}, C.~J., N.~A. {Schwadron}, L.~W. {Townsend}, R.~A. {Mewaldt}, C.~M.~S.
  {Cohen}, T.~T. {Rosenvinge}, A.~W. {Case}, H.~E. {Spence}, J.~K. {Wilson},
  M.~{Gorby}, M.~{Quinn}, and C.~J. {Zeitlin},
\newblock {Analysis of the potential radiation hazard of the 23 July 2012 SEP
  event observed by STEREO A using the EMMREM model and LRO/CRaTER},
\newblock {\em Space Weather}, {\em 13}, 560--567, September 2015.

\bibitem[{\em {Kahler} et~al.}(1978)]{kahler78}
{Kahler}, S.~W., E.~{Hildner}, and M.~A.~I. {Van Hollebeke},
\newblock {Prompt solar proton events and coronal mass ejections},
\newblock {\em \solphys}, {\em 57}, 429--443, April 1978.

\bibitem[{\em {Kahler} et~al.}(2012)]{kahler12}
{Kahler}, S.~W., E.~W. {Cliver}, A.~J. {Tylka}, and W.~F. {Dietrich},
\newblock {A Comparison of Ground Level Event e/p and Fe/O Ratios with
  Associated Solar Flare and CME Characteristics},
\newblock {\em \ssr}, {\em 171}, 121--139, October 2012.

\bibitem[{\em {Kahler}}(1994)]{Kahler1994}
{Kahler}, S.,
\newblock {Injection profiles of solar energetic particles as functions of
  coronal mass ejection heights},
\newblock {\em \apj}, {\em 428}, 837--842, June 1994.

\bibitem[{\em {K{\'o}ta} et~al.}(2005)]{Kota2005}
{K{\'o}ta}, J., W.~B. {Manchester}, J.~R. {Jokipii}, D.~L. {de Zeeuw}, and
  T.~I. {Gombosi},
\newblock {Simulation of SEP Acceleration and Transport at CME-driven Shocks},
\newblock In {Li}, G., G.~P. {Zank}, and C.~T. {Russell}, editors, {\em The
  Physics of Collisionless Shocks: 4th Annual IGPP International Astrophysics
  Conference}, volume 781 of {\em American Institute of Physics Conference
  Series}, pages 201--206, August 2005.

\bibitem[{\em {Krymsky}}(1977)]{Krymsky1977}
{Krymsky}, G.~F.,
\newblock {A regular mechanism for the acceleration of charged particles on the
  front of a shock wave},
\newblock {\em Akademiia Nauk SSSR Doklady}, {\em 234}, 1306--1308, June 1977.

\bibitem[{\em {Landau} and {Lifshitz}}(1960)]{landau60}
{Landau}, L.~D., and E.~M. {Lifshitz},
\newblock {\em {Electrodynamics of continuous media}},
\newblock Pergamon Press: Oxford, 1960.

\bibitem[{\em {Landi} et~al.}(2013)]{landi13}
{Landi}, E., P.~R. {Young}, K.~P. {Dere}, G.~{Del Zanna}, and H.~E. {Mason},
\newblock {CHIANTI - An Atomic Database for Emission Lines. XIII. Soft X-Ray
  Improvements and Other Changes},
\newblock {\em \apj}, {\em 763}, 86, February 2013.

\bibitem[{\em {Lee}}(1997)]{Lee1997}
{Lee}, M.~A.,
\newblock {Particle Acceleration and Transport at CME-Driven Shocks, in Coronal
  Mass Ejections (eds N. Crooker, J. A. Joselyn and J. Feynman)},
\newblock {\em Washington DC American Geophysical Union Geophysical Monograph
  Series}, {\em 99}, 227--234, 1997.

\bibitem[{\em {Leroy}}(1980)]{Leroy1980}
{Leroy}, B.,
\newblock {Propagation of waves in an atmosphere in the presence of a magnetic
  field. II - The reflection of Alfven waves},
\newblock {\em \aap}, {\em 91}, 136--146, November 1980.

\bibitem[{\em {Li} et~al.}(2003)]{li03}
{Li}, G., G.~P. {Zank}, and W.~K.~M. {Rice},
\newblock {Energetic Particle Acceleration and Transport at Coronal Mass
  Ejection-Driven Shocks},
\newblock {\em J. Geophys. Res.}, {\em 108(A2)}, 10--21, February 2003.

\bibitem[{\em {Linker} et~al.}(2016)]{Linker2016}
{Linker}, J., T.~{Torok}, C.~{Downs}, R.~{Lionello}, V.~{Titov}, R.~M.
  {Caplan}, Z.~{Miki{\'c}}, and P.~{Riley},
\newblock {MHD simulation of the Bastille day event},
\newblock In {\em American Institute of Physics Conference Series}, volume 1720
  of {\em American Institute of Physics Conference Series}, page 020002, March
  2016.

\bibitem[{\em {Lionello} et~al.}(2001)]{Lionello2001}
{Lionello}, R., J.~A. {Linker}, and Z.~{Miki{\'c}},
\newblock {Including the Transition Region in Models of the Large-Scale Solar
  Corona},
\newblock {\em \apj}, {\em 546}, 542--551, January 2001.

\bibitem[{\em {Lionello} et~al.}(2009)]{Lionello2009}
{Lionello}, R., J.~A. {Linker}, and Z.~{Miki{\'c}},
\newblock {Multispectral Emission of the Sun During the First Whole Sun Month:
  Magnetohydrodynamic Simulations},
\newblock {\em \apj}, {\em 690}, 902--912, January 2009.

\bibitem[{\em {Lionello} et~al.}(2014a)]{Lionello2014a}
{Lionello}, R., M.~{Velli}, C.~{Downs}, J.~A. {Linker}, and Z.~{Miki{\'c}},
\newblock {Application of a Solar Wind Model Driven by Turbulence Dissipation
  to a 2D Magnetic Field Configuration},
\newblock {\em \apj}, {\em 796}, 111, December 2014a.

\bibitem[{\em {Lionello} et~al.}(2014b)]{Lionello2014b}
{Lionello}, R., M.~{Velli}, C.~{Downs}, J.~A. {Linker}, Z.~{Miki{\'c}}, and
  A.~{Verdini},
\newblock {Validating a Time-dependent Turbulence-driven Model of the Solar
  Wind},
\newblock {\em \apj}, {\em 784}, 120, April 2014b.

\bibitem[{\em {Low}}(1982)]{low82}
{Low}, B.~C.,
\newblock {Self-similar magnetohydrodynamics. I - The gamma = 4/3 polytrope and
  the coronal transient},
\newblock {\em \apj}, {\em 254}, 796--805, March 1982.

\bibitem[{\em {Lugaz} et~al.}(2005)]{lugaz05}
{Lugaz}, N., W.~B. {Manchester}, IV, and T.~I. {Gombosi},
\newblock Numerical simulation of the interaction of two coronal mass ejections
  from sun to earth,
\newblock {\em \apj}, {\em 634}, 651--662, November 2005.

\bibitem[{\em {Lugaz} et~al.}(2007)]{Lugaz2007}
{Lugaz}, N., W.~B. {Manchester}, IV, I.~I. {Roussev}, G.~{T{\'o}th}, and T.~I.
  {Gombosi},
\newblock {Numerical Investigation of the Homologous Coronal Mass Ejection
  Events from Active Region 9236},
\newblock {\em \apj}, {\em 659}, 788--800, April 2007.

\bibitem[{\em {MacNeice}}(2009)]{MacNeice2009}
{MacNeice}, P.,
\newblock {Validation of community models: 2. Development of a baseline using
  the Wang-Sheeley-Arge model},
\newblock {\em Space Weather}, {\em 7}, S12002, December 2009.

\bibitem[{\em Makhmutov et~al.}(2009)]{Makhmutov:2009}
Makhmutov, V.S., G.A. Bazilevskaya, Y.I. Stozhkov, N.S. Svirzhevsky, and A.K.
  Svirzhevskaya,
\newblock Ionisation state of the {Earth}'s stratosphere during powerful solar
  proton events,
\newblock In {\em Proc. 31st International Cosmic Ray Conference},
  \L{}{\'o}d{\'z}, Poland, 2009.

\bibitem[{\em {Manchester} et~al.}(2004a)]{manchester04a}
{Manchester}, W.~B., T.~I. {Gombosi}, I.~{Roussev}, D.~L. {De Zeeuw}, I.~V.
  {Sokolov}, K.~G. {Powell}, G.~{T{\'o}th}, and M.~{Opher},
\newblock Three-dimensional {MHD} simulation of a flux rope driven {CME},
\newblock {\em J. Geophys. Res.}, {\em 109(A18)}, 1,102--1,119, January 2004a.

\bibitem[{\em {Manchester} et~al.}(2004b)]{manchester04b}
{Manchester}, W.~B., T.~I. {Gombosi}, I.~{Roussev}, A.~{Ridley}, D.~L. {De
  Zeeuw}, I.~V. {Sokolov}, K.~G. {Powell}, and G.~{T{\'o}th},
\newblock Modeling a space weather event from the {Sun} to the {Earth}: {CME}
  generation and interplanetary propagation,
\newblock {\em J. Geophys. Res.}, {\em 109(A18)}, 2,107--2,122, February 2004b.

\bibitem[{\em Manchester et~al.}(2005)]{Manchester2005}
Manchester, W.~B., T.~I. Gombosi, D.~L. {De~Zeeuw}, I.~V. Sokolov, I.~I.
  Roussev, K.~G. Powell, J.~K\'ota, G.~T\'oth, and T.~H. Zurbuchen,
\newblock Coronal mass ejection shock and sheath structures relevant to
  particle acceleration,
\newblock {\em AStrophys. J.}, {\em 622}, 1225--1239, 2005.

\bibitem[{\em {Manchester} et~al.}(2006)]{manchester06}
{Manchester}, W.~B., A.~J. {Ridley}, T.~I. {Gombosi}, and D.~L. {DeZeeuw},
\newblock {Modeling the Sun-to-Earth propagation of a very fast CME},
\newblock {\em Advances in Space Research}, {\em 38}, 253--262, January 2006.

\bibitem[{\em {Manchester} et~al.}(2008)]{manchester08}
{Manchester}, W.~B., IV, A.~{Vourlidas}, G.~{T{\'o}th}, N.~{Lugaz}, I.~I.
  {Roussev}, I.~V. {Sokolov}, T.~I. {Gombosi}, D.~L. {De Zeeuw}, and
  M.~{Opher},
\newblock {Three-dimensional MHD Simulation of the 2003 October 28 Coronal Mass
  Ejection: Comparison with LASCO Coronagraph Observations},
\newblock {\em \apj}, {\em 684}, 1448--1460, September 2008.

\bibitem[{\em {Manchester} et~al.}(2012)]{Manchester2012}
{Manchester}, W.~B., IV, B.~{van der Holst}, G.~{T{\'o}th}, and T.~I.
  {Gombosi},
\newblock {The Coupled Evolution of Electrons and Ions in Coronal Mass
  Ejection-driven shocks},
\newblock {\em \apj}, {\em 756}, 81, September 2012.

\bibitem[{\em {Matsumoto} and {Suzuki}}(2012)]{Matsumoto2012}
{Matsumoto}, T., and T.~K. {Suzuki},
\newblock {Connecting the Sun and the Solar Wind: The First 2.5-dimensional
  Self-consistent MHD Simulation under the Alfv{\'e}n Wave Scenario},
\newblock {\em \apj}, {\em 749}, 8, April 2012.

\bibitem[{\em {Matthaeus} et~al.}(1999)]{Matthaeus1999}
{Matthaeus}, W.~H., G.~P. {Zank}, S.~{Oughton}, D.~J. {Mullan}, and
  P.~{Dmitruk},
\newblock {Coronal Heating by Magnetohydrodynamic Turbulence Driven by
  Reflected Low-Frequency Waves},
\newblock {\em \apjl}, {\em 523}, L93--L96, September 1999.

\bibitem[{\em {Matthi{\"a}} et~al.}(2009)]{matthia09}
{Matthi{\"a}}, D., B.~{Heber}, G.~{Reitz}, M.~{Meier}, L.~{Sihver},
  T.~{Berger}, and K.~{Herbst},
\newblock {Temporal and spatial evolution of the solar energetic particle event
  on 20 January 2005 and resulting radiation doses in aviation},
\newblock {\em Journal of Geophysical Research (Space Physics)}, {\em 114},
  A08104, August 2009.

\bibitem[{\em {Mays} et~al.}(2015)]{mays15}
{Mays}, M.~L., A.~{Taktakishvili}, A.~{Pulkkinen}, P.~J. {MacNeice},
  L.~{Rast{\"a}tter}, D.~{Odstrcil}, L.~K. {Jian}, I.~G. {Richardson}, J.~A.
  {LaSota}, Y.~{Zheng}, and M.~M. {Kuznetsova},
\newblock {Ensemble Modeling of CMEs Using the WSA-ENLIL+Cone Model},
\newblock {\em \solphys}, {\em 290}, 1775--1814, June 2015.

\bibitem[{\em {Mazur} et~al.}(1999)]{mazur99}
{Mazur}, J.~E., G.~M. {Mason}, M.~D. {Looper}, R.~A. {Leske}, and R.~A.
  {Mewaldt},
\newblock {Charge states of solar energetic particles using the geomagnetic
  cutoff technique: SAMPEX measurements in the 6 November 1997 solar particle
  event},
\newblock {\em \grl}, {\em 26}, 173--176, 1999.

\bibitem[{\em {Meyer} et~al.}(1956)]{meyer56}
{Meyer}, P., E.~N. {Parker}, and J.~A. {Simpson},
\newblock {Solar Cosmic Rays of February, 1956 and Their Propagation through
  Interplanetary Space},
\newblock {\em Physical Review}, {\em 104}, 768--783, November 1956.

\bibitem[{\em {Michalek}}(2006)]{Michalek2006}
{Michalek}, G.,
\newblock {An Asymmetric Cone Model for Halo Coronal Mass Ejections},
\newblock {\em \solphys}, {\em 237}, 101--118, August 2006.

\bibitem[{\em {Miki\'c} et~al.}(2007)]{Mikic2007}
{Miki\'c}, Z., J.A. {Linker}, R.~{Lionello}, P.~{Riley}, and V.~{Titov},
\newblock Predicting the structure of the solar corona for the total solar
  eclipse of march 29, 2006,
\newblock {\em Solar and stellar physics through eclipses, {ASP} Conf. Ser.},
  {\em 370}, 299--307, 2007.

\bibitem[{\em {Morris}}(2007)]{Morris07}
{Morris}, Doug,
\newblock {\em From the Flight Deck: {P}lane Talk and Sky Science},
\newblock ECW Press, Toronto, Ontario, Canada, 2007.

\bibitem[{\em {Nesse Tyss{\o}y} et~al.}(2013)]{tyssoy13}
{Nesse Tyss{\o}y}, H., J.~{Stadsnes}, F.~{S{\o}Raas}, and M.~{S{\o}Rb{\o}},
\newblock {Variations in cutoff latitude during the January 2012 solar proton
  event and implication for the distribution of particle energy deposition},
\newblock {\em \grl}, {\em 40}, 4149--4153, August 2013.

\bibitem[{\em {Ng} et~al.}(1999)]{Ng1999}
{Ng}, C.~K., D.~V. {Reames}, and A.~J. {Tylka},
\newblock {Effect of proton-amplified waves on the evolution of solar energetic
  particle composition in gradual events},
\newblock {\em \grl}, {\em 26}, 2145--2148, 1999.

\bibitem[{\em {Ng} et~al.}(2003)]{Ng2003}
{Ng}, C.~K., D.~V. {Reames}, and A.~J. {Tylka},
\newblock {Modeling Shock-accelerated Solar Energetic Particles Coupled to
  Interplanetary Alfv{\'e}n Waves},
\newblock {\em \apj}, {\em 591}, 461--485, July 2003.

\bibitem[{\em {Norquist} and {Meeks}}(2010)]{Norquist2010}
{Norquist}, D.~C., and W.~C. {Meeks},
\newblock {A comparative verification of forecasts from two operational solar
  wind models},
\newblock {\em Space Weather}, {\em 8}, S12005, December 2010.

\bibitem[{\em {Oran} et~al.}(2013)]{Oran2013}
{Oran}, R., B.~{van der Holst}, E.~{Landi}, M.~{Jin}, I.~V. {Sokolov}, and
  T.~I. {Gombosi},
\newblock {A Global Wave-driven Magnetohydrodynamic Solar Model with a Unified
  Treatment of Open and Closed Magnetic Field Topologies},
\newblock {\em \apj}, {\em 778}, 176, December 2013.

\bibitem[{\em {Owens} et~al.}(2008)]{Owens2008}
{Owens}, M.~J., H.~E. {Spence}, S.~{McGregor}, W.J. {Hughes}, J.~M. {Quinn},
  C.~N. {Arge}, P.~{Riley}, J.~{Linker}, and D.~Odstr{\v c}il,
\newblock Metrics for solar wind prediction models: Comparison of empirical,
  hybrid and physics-based schemes with 8-years of l1 observations,
\newblock {\em Space Weather}, {\em 6}, 2008.

\bibitem[{\em {Parker}}(1965)]{Parker1965}
{Parker}, E.~N.,
\newblock {The passage of energetic charged particles through interplanetary
  space},
\newblock {\em \planss}, {\em 13}, 9--49, January 1965.

\bibitem[{\em {Plunkett} et~al.}(1998)]{plunkett98}
{Plunkett}, S.~P., B.~J. {Thompson}, R.~A. {Howard}, D.~J. {Michels}, O.~C.
  {St.~Cyr}, S.~J. {Tappin}, R.~{Schwenn}, and P.~L. {Lamy},
\newblock {LASCO observations of an Earth-directed coronal mass ejection on May
  12, 1997},
\newblock {\em \grl}, {\em 25}, 2477--2480, 1998.

\bibitem[{\em {Reames}}(1999)]{reames99}
{Reames}, D.~V.,
\newblock {Particle Acceleration at the Sun and in the Heliosphere},
\newblock {\em Space Sci. Rev.}, {\em 90}, 413--491, February 1999.

\bibitem[{\em {Reames}}(2002)]{Reames2002}
{Reames}, D.~V.,
\newblock {Magnetic Topology of Impulsive and Gradual Solar Energetic Particle
  Events},
\newblock {\em \apjl}, {\em 571}, L63--L66, May 2002.

\bibitem[{\em {Reiss} et~al.}(2016)]{Reiss2016}
{Reiss}, M.~A., M.~{Temmer}, A.~M. {Veronig}, L.~{Nikolic}, S.~{Vennerstrom},
  F.~{Sch{\"o}ngassner}, and S.~J. {Hofmeister},
\newblock {Verification of high-speed solar wind stream forecasts using
  operational solar wind models},
\newblock {\em Space Weather}, {\em 14}, 495--510, July 2016.

\bibitem[{\em {Rice} et~al.}(2003)]{Rice2003}
{Rice}, W.~K.~M., G.~P. {Zank}, and G.~{Li},
\newblock {Particle acceleration and coronal mass ejection driven shocks:
  Shocks of arbitrary strength},
\newblock {\em Journal of Geophysical Research (Space Physics)}, {\em 108},
  1369, October 2003.

\bibitem[{\em {Riley} et~al.}(2006)]{Riley2006}
{Riley}, P., J.~A. {Linker}, Z.~{Miki{\'c}}, R.~{Lionello}, S.~A. {Ledvina},
  and J.~G. {Luhmann},
\newblock {A Comparison between Global Solar Magnetohydrodynamic and Potential
  Field Source Surface Model Results},
\newblock {\em \apj}, {\em 653}, 1510--1516, December 2006.

\bibitem[{\em {Roussev} and {Sokolov}}(2006)]{Roussev2006}
{Roussev}, I.~I., and I.~V. {Sokolov},
\newblock {Models of Solar Eruptions: Recent Advances from Theory and
  Simulations, in Solar Eruptions and Energetic Particles (eds N. Gopalswamy,
  R. Mewaldt and J. Torsti)},
\newblock {\em Washington DC American Geophysical Union Geophysical Monograph
  Series}, {\em 165}, October 2006.

\bibitem[{\em {Roussev} et~al.}(2003a)]{Roussev2003a}
{Roussev}, I.~I., T.~G. {Forbes}, T.~I. {Gombosi}, I.~V. {Sokolov}, D.~L.
  {DeZeeuw}, and J.~{Birn},
\newblock {A Three-dimensional Flux Rope Model for Coronal Mass Ejections Based
  on a Loss of Equilibrium},
\newblock {\em \apjl}, {\em 588}, L45--L48, May 2003a.

\bibitem[{\em {Roussev} et~al.}(2003b)]{Roussev2003}
{Roussev}, I.~I., T.~I. {Gombosi}, I.~V. {Sokolov}, M.~{Velli},
  W.~{Manchester}, IV, D.~L. {DeZeeuw}, P.~{Liewer}, G.~{T{\'o}th}, and
  J.~{Luhmann},
\newblock {A Three-dimensional Model of the Solar Wind Incorporating Solar
  Magnetogram Observations},
\newblock {\em \apjl}, {\em 595}, L57--L61, September 2003b.

\bibitem[{\em {Roussev} et~al.}(2004)]{Roussev2004}
{Roussev}, I.~I., I.~V. {Sokolov}, T.~G. {Forbes}, T.~I. {Gombosi}, M.~A.
  {Lee}, and J.~I. {Sakai},
\newblock {A Numerical Model of a Coronal Mass Ejection: Shock Development with
  Implications for the Acceleration of GeV Protons},
\newblock {\em \apjl}, {\em 605}, L73--L76, April 2004.

\bibitem[{\em {Ruffolo} et~al.}(1998)]{Ruffolo1998}
{Ruffolo}, D., T.~{Khumlumlert}, and W.~{Youngdee},
\newblock {Deconvolution of interplanetary transport of solar energetic
  particles},
\newblock {\em \jgr}, {\em 103}, 20591--20602, September 1998.

\bibitem[{\em {Sedov}}(1959)]{sedov59}
{Sedov}, L.~I.,
\newblock {\em {Similarity and Dimensional Methods in Mechanics}},
\newblock 1959.

\bibitem[{\em {Shafranov}}(1966)]{shafranov66}
{Shafranov}, V.~D.,
\newblock {Plasma Equilibrium in a Magnetic Field},
\newblock {\em Rev. Plasma Phys.}, {\em 2}, 103, 1966.

\bibitem[{\em {Shea} and {Smart}}(1990)]{Shea:1990}
{Shea}, M.~A., and D.~F. {Smart},
\newblock A summary of major solar proton events,
\newblock {\em Solar Physics}, {\em 127}( 2 ), 297--320, 1990.

\bibitem[{\em {Shea} and {Smart}}(1994)]{shea94}
{Shea}, M.~A., and D.~F. {Smart},
\newblock {Significant proton events of solar cycle 22 and a comparison with
  events of previous solar cycles},
\newblock {\em Advances in Space Research}, {\em 14}, 631--638, October 1994.

\bibitem[{\em {Shea} and Smart}(2006)]{Shea06a}
{Shea}, M.~A., and D.~F. Smart,
\newblock Compendium of the eight articles on the ``{C}arrington {E}vent''
  attributed to or written by {Elias Loomis} in the {American Journal of
  Science}, 1859-1861,
\newblock {\em Adv. Space Res.}, {\em 38}( 2 ), 313--385, 2006.

\bibitem[{\em {Shea} and {Smart}}(2012)]{shea12}
{Shea}, M.~A., and D.~F. {Smart},
\newblock {Space Weather and the Ground-Level Solar Proton Events of the 23rd
  Solar Cycle},
\newblock {\em \ssr}, {\em 171}, 161--188, October 2012.

\bibitem[{\em {Shea} et~al.}(2006)]{Shea06}
{Shea}, M.~A., D.F. {Smart}, K.~G. {McCracken}, G.~A.~M. {Dreschhoff}, and H.E.
  {Spence},
\newblock Solar proton events for 450 years: {T}he {C}arrington event in
  perspective,
\newblock {\em Adv. Space Res.}, {\em 38}( 2 ), 232--238, 2006.

\bibitem[{\em {Sokolov} et~al.}(2004)]{Sokolov2004}
{Sokolov}, I.~V., I.~I. {Roussev}, T.~I. {Gombosi}, M.~A. {Lee}, J.~{K{\'o}ta},
  T.~G. {Forbes}, W.~B. {Manchester}, and J.~I. {Sakai},
\newblock {A New Field Line Advection Model for Solar Particle Acceleration},
\newblock {\em \apjl}, {\em 616}, L171--L174, December 2004.

\bibitem[{\em {Sokolov} et~al.}(2009)]{Sokolov2009}
{Sokolov}, I.~V., I.~I. {Roussev}, M.~{Skender}, T.~I. {Gombosi}, and A.~V.
  {Usmanov},
\newblock {Transport Equation for MHD Turbulence: Application to Particle
  Acceleration at Interplanetary Shocks},
\newblock {\em \apj}, {\em 696}, 261--267, May 2009.

\bibitem[{\em {Sokolov} et~al.}(2013)]{Sokolov2013}
{Sokolov}, I.~V., B.~{van der Holst}, R.~{Oran}, C.~{Downs}, I.~I. {Roussev},
  M.~{Jin}, W.~B. {Manchester}, IV, R.~M. {Evans}, and T.~I. {Gombosi},
\newblock {Magnetohydrodynamic Waves and Coronal Heating: Unifying Empirical
  and MHD Turbulence Models},
\newblock {\em \apj}, {\em 764}, 23, February 2013.

\bibitem[{\em {Sokolov} et~al.}(2016)]{sokolov16}
{Sokolov}, I.~V., B.~{van der Holst}, W.~B. {Manchester}, D.~C.~S. {Ozturk},
  J.~{Szente}, A.~{Taktakishvili}, G.~{T{\'o}th}, M.~{Jin}, and T.~I.
  {Gombosi},
\newblock {Threaded-Field-Lines Model for the Low Solar Corona Powered by the
  Alfven Wave Turbulence},
\newblock {\em ArXiv e-prints}, September 2016.

\bibitem[{\em {Spitzer} and {H{\"a}rm}}(1953)]{spitzer53}
{Spitzer}, L., and R.~{H{\"a}rm},
\newblock {Transport Phenomena in a Completely Ionized Gas},
\newblock {\em Physical Review}, {\em 89}, 977--981, March 1953.

\bibitem[{\em {Suzuki} and {Inutsuka}}(2005)]{Suzuki2005}
{Suzuki}, T.~K., and S.-i. {Inutsuka},
\newblock {Making the Corona and the Fast Solar Wind: A Self-consistent
  Simulation for the Low-Frequency Alfv{\'e}n Waves from the Photosphere to 0.3
  AU},
\newblock {\em \apjl}, {\em 632}, L49--L52, October 2005.

\bibitem[{\em {Titov} et~al.}(2008)]{Titov2008}
{Titov}, V.~S., Z.~{Mikic}, J.~A. {Linker}, and R.~{Lionello},
\newblock {1997 May 12 Coronal Mass Ejection Event. I. A Simplified Model of
  the Preeruptive Magnetic Structure},
\newblock {\em \apj}, {\em 675}, 1614--1628, March 2008.

\bibitem[{\em {Titov} et~al.}(2014)]{titov14}
{Titov}, V.~S., T.~{T{\"o}r{\"o}k}, Z.~{Mikic}, and J.~A. {Linker},
\newblock {A Method for Embedding Circular Force-free Flux Ropes in Potential
  Magnetic Fields},
\newblock {\em \apj}, {\em 790}, 163, August 2014.

\bibitem[{\em {T{\"o}r{\"o}k} and {Kliem}}(2005)]{Torok2005}
{T{\"o}r{\"o}k}, T., and B.~{Kliem},
\newblock {Confined and Ejective Eruptions of Kink-unstable Flux Ropes},
\newblock {\em \apjl}, {\em 630}, L97--L100, September 2005.

\bibitem[{\em {T{\'o}th} et~al.}(2007)]{Toth2007}
{T{\'o}th}, G., Darren~L. De~Zeeuw, Tamas~I. Gombosi, Ward~B. Manchester,
  Aaron~J. Ridley, Igor~V. Sokolov, and Ilia~I. Roussev,
\newblock Sun-to-thermosphere simulation of the 28--30 october 2003 storm with
  the space weather modeling framework,
\newblock {\em Space Weather}, {\em 5}( 6 ), n/a--n/a, 2007,
\newblock S06003.

\bibitem[{\em {T{\'o}th} et~al.}(2012)]{Toth2012}
{T{\'o}th}, G., B.~{van der Holst}, I.~V. {Sokolov}, D.~L. {De Zeeuw}, T.~I.
  {Gombosi}, F.~{Fang}, W.~B. {Manchester}, X.~{Meng}, D.~{Najib}, K.~G.
  {Powell}, Q.~F. {Stout}, A.~{Glocer}, Y.-J. {Ma}, and M.~{Opher},
\newblock {Adaptive numerical algorithms in space weather modeling},
\newblock {\em Journal of Computational Physics}, {\em 231}, 870--903, February
  2012.

\bibitem[{\em {Tylka} et~al.}(1999)]{Tylka1999}
{Tylka}, A.~J., D.~V. {Reames}, and C.~K. {Ng},
\newblock {Observations of systematic temporal evolution in elemental
  composition during gradual solar energetic particle events},
\newblock {\em \grl}, {\em 26}, 2141--2144, 1999.

\bibitem[{\em {Tylka} et~al.}(2005)]{tylka05}
{Tylka}, A.~J., C.~M.~S. {Cohen}, W.~F. {Dietrich}, M.~A. {Lee}, C.~G.
  {Maclennan}, R.~A. {Mewaldt}, C.~K. {Ng}, and D.~V. {Reames},
\newblock {Shock Geometry, Seed Populations, and the Origin of Variable
  Elemental Composition at High Energies in Large Gradual Solar Particle
  Events},
\newblock {\em Astrophys. J.}, {\em 625}, 474--495, May 2005.

\bibitem[{\em {Tylka}}(2001)]{Tylka2001}
{Tylka}, A.~J.,
\newblock {New insights on solar energetic particles from Wind and ACE},
\newblock {\em \jgr}, {\em 106}, 25333--25352, November 2001.

\bibitem[{\em {Usmanov} et~al.}(2000)]{Usmanov2000}
{Usmanov}, A.~V., M.~L. {Goldstein}, B.~P. {Besser}, and J.~M. {Fritzer},
\newblock {A global MHD solar wind model with WKB Alfv{\'e}n waves: Comparison
  with Ulysses data},
\newblock {\em \jgr}, {\em 105}, 12675--12696, June 2000.

\bibitem[{\em {van der Holst} et~al.}(2014)]{Holst2014}
{van der Holst}, B., I.~V. {Sokolov}, X.~{Meng}, M.~{Jin}, W.~B. {Manchester},
  IV, G.~{T{\'o}th}, and T.~I. {Gombosi},
\newblock {Alfv{\'e}n Wave Solar Model (AWSoM): Coronal Heating},
\newblock {\em Astrophys J.}, {\em 782}, 81, February 2014.

\bibitem[{\em {V{\'a}squez} et~al.}(2008)]{Vasquez2008}
{V{\'a}squez}, A.~M., R.~A. {Frazin}, K.~{Hayashi}, I.~V. {Sokolov},
  O.~{Cohen}, W.~B. {Manchester}, IV, and F.~{Kamalabadi},
\newblock {Validation of Two MHD Models of the Solar Corona with Rotational
  Tomography},
\newblock {\em \apj}, {\em 682}, 1328--1337, August 2008.

\bibitem[{\em {Verdini} and {Velli}}(2007)]{Verdini2007}
{Verdini}, A., and M.~{Velli},
\newblock {Alfv{\'e}n Waves and Turbulence in the Solar Atmosphere and Solar
  Wind},
\newblock {\em \apj}, {\em 662}, 669--676, June 2007.

\bibitem[{\em {Verdini} et~al.}(2010)]{Verdini2010}
{Verdini}, A., M.~{Velli}, W.~H. {Matthaeus}, S.~{Oughton}, and P.~{Dmitruk},
\newblock {A Turbulence-Driven Model for Heating and Acceleration of the Fast
  Wind in Coronal Holes},
\newblock {\em \apjl}, {\em 708}, L116--L120, January 2010.

\bibitem[{\em {Vr{\v s}nak} et~al.}(2014)]{vrsnak14}
{Vr{\v s}nak}, B., M.~{Temmer}, T.~{{\v Z}ic}, A.~{Taktakishvili},
  M.~{Dumbovi{\'c}}, C.~{M{\"o}stl}, A.~M. {Veronig}, M.~L. {Mays}, and
  D.~{Odstr{\v c}il},
\newblock {Heliospheric Propagation of Coronal Mass Ejections: Comparison of
  Numerical WSA-ENLIL+Cone Model and Analytical Drag-based Model},
\newblock {\em \apjs}, {\em 213}, 21, August 2014.

\bibitem[{\em {Wang} and {Sheeley}}(1990)]{Wang1990}
{Wang}, Y.-M., and N.~R. {Sheeley}, Jr.,
\newblock {Solar wind speed and coronal flux-tube expansion},
\newblock {\em \apj}, {\em 355}, 726--732, June 1990.

\bibitem[{\em {Wang} and {Sheeley}}(1992)]{Wang1992}
{Wang}, Y.-M., and N.~R. {Sheeley}, Jr.,
\newblock {On potential field models of the solar corona},
\newblock {\em \apj}, {\em 392}, 310--319, June 1992.

\bibitem[{\em {Wang} and {Sheeley}}(1995)]{Wang1995}
{Wang}, Y.-M., and N.~R. {Sheeley}, Jr.,
\newblock {Solar Implications of ULYSSES Interplanetary Field Measurements},
\newblock {\em \apjl}, {\em 447}, L143, July 1995.

\bibitem[{\em {Wild} et~al.}(1963)]{wild63}
{Wild}, J.~P., S.~F. {Smerd}, and A.~A. {Weiss},
\newblock {Solar Bursts},
\newblock {\em \araa}, {\em 1}, 291, 1963.

\bibitem[{\em {Zank} et~al.}(2000)]{Zank2000}
{Zank}, G.~P., W.~K.~M. {Rice}, and C.~C. {Wu},
\newblock {Particle acceleration and coronal mass ejection driven shocks: A
  theoretical model},
\newblock {\em \jgr}, {\em 105}, 25079--25096, November 2000.

\bibitem[{\em {Zel'dovich} and {Raizer}}(1967)]{zeldovich67}
{Zel'dovich}, Y.~B., and Y.~P. {Raizer},
\newblock {\em {Physics of shock waves and high-temperature hydrodynamic
  phenomena}},
\newblock 1967.

\bibitem[{\em {Zhao} et~al.}(2002)]{Zhao2002}
{Zhao}, X.~P., S.~P. {Plunkett}, and W.~{Liu},
\newblock {Determination of geometrical and kinematical properties of halo
  coronal mass ejections using the cone model},
\newblock {\em Journal of Geophysical Research (Space Physics)}, {\em 107},
  1223, August 2002.

\end{thebibliography}
\clearpage
\end{document}